\documentclass{article}

\usepackage{arxiv}

\usepackage[utf8]{inputenc} 
\usepackage[T1]{fontenc}    
\usepackage{hyperref}       
\usepackage{url}            
\usepackage{booktabs}       
\usepackage{amsfonts}       
\usepackage{nicefrac}       
\usepackage{microtype}      
\usepackage{amsmath}
\usepackage{amsfonts,amssymb}
\usepackage{bm}
\usepackage{multirow}
\usepackage{graphicx}
\usepackage{epstopdf}
\usepackage{subfigure}
\usepackage{parskip}
\usepackage{color,xcolor}
\usepackage{hyperref}
\usepackage[section]{placeins}

\newtheorem{corollary}{Corollary}

\newtheorem{definition}{Definition}

\newcommand{\corref}[1]{Corollary~\ref{#1}}

\newcommand{\y}[1]{\textcolor{black}{#1}}
\def\sph{\mathbb{S}^{2}}
\def\RR{\mathbb{R}}

\def\CB{\mathcal{B}}

\def\NN{\mathbb{N}}
\def\f{\frac}

\def\b{\bm}
\def\var{\varepsilon}
\def\h{\mathcal{F}}

\title{Spherical Image Inpainting with Frame Transformation and Data-driven Prior Deep Networks}
\author{ {\bf Jianfei Li}\thanks{Co-first authors. Department of Mathematics, City University of Hong Kong
		(\texttt{jianfeili2-c@my.cityu.edu.hk})}
	\and {\bf Chaoyan Huang}\thanks{Co-first authors. Department of Mathematics, The Chinese University of Hong Kong
		(\texttt{cyhuang@math.cuhk.edu.hk})}
	\and {\bf Raymond Chan}\thanks{Department of Mathematics, City University of Hong Kong and Hong Kong Centre for Cerebro-Cardiovascular Health Engineering (\texttt{raymond.chan@cityu.edu.hk})}
	\and {\bf Han Feng}\thanks{Corresponding author. Department of Mathematics, City University of Hong Kong (\texttt{hanfeng@cityu.edu.hk})}
	\and {\bf Micheal Ng}\thanks{Department of Mathematics, University of Hong Kong (\texttt{mng@maths.hku.hk})}
	\and {\bf Tieyong Zeng}\thanks{Department of Mathematics, The Chinese University of Hong Kong (\texttt{zeng@math.cuhk.edu.hk})}
}

\begin{document}
\maketitle
\begin{abstract}
Spherical image processing has been widely applied in many important fields, such as omnidirectional vision for autonomous cars, global climate modelling, and medical imaging. It is non-trivial to extend an algorithm developed for flat images to the spherical ones. In this work, we focus on the challenging task of spherical image inpainting with deep learning-based regularizer. Instead of a naive application of existing models for planar images, we employ a fast directional spherical Haar framelet transform and develop a novel optimization framework based on a sparsity assumption of the framelet transform. Furthermore, by employing progressive encoder-decoder architecture, a new and better-performed deep CNN denoiser is carefully designed and works as an implicit regularizer. Finally, we use a plug-and-play method to handle the proposed optimization model, which can be implemented efficiently by training the CNN denoiser prior. Numerical experiments are conducted and show that the proposed algorithms can greatly recover damaged spherical images and achieve the best performance over purely using deep learning denoiser and plug-and-play model.

\bf{\emph{ Key words-\ Spherical image inpainting; deep CNN; plug and play}\rm}
\end{abstract}



\section{Introduction}

In practical problems, a large amount of data comes in the form of spherical images, such as  from cosmology
\cite{cosmology}, astrophysics \cite{astrophysics}, geophysics \cite{planetary,geo}, neuroscience
\cite{neuroscience}, and omnidirectional AR/VR field \cite{omni,omni2}, where images are naturally defined on the 2D spherical surface. 
Due to the storage bottleneck and observation being costly and infeasible, these spherical images (signals) usually contain very limited pixels (observed data), especially if the observation scales involved are large. Therefore, repairing missing or damaged parts is a fundamental yet challenging task in spherical image processing.
Apparently, spherical images take a different inherent domain than planar images in 2D in terms of symmetries, coordinate systems, and translates, which demand special processing methods.
In this paper, we are concerned with spherical image restoration, which can further serve as a preliminary for 
subsequent tasks, like object recognition and segmentation. Mathematically, it aims to estimate $x$ from observation $y$ for the following model
 \begin{equation}\label{degraded}
y=T(x)+\var,
\end{equation}
where $T$ is a degradation operator, $\var$ is assumed to be the additive noise. Different degradation operations correspond to different image restoration (IR) tasks \cite{jia2021structure,ke2020reconstruction}.
Typically, the IR task would be image denoising when $T$ is an identity operation,
image deblurring when $T$ is a two-dimensional convolution
operation, image super-resolution when $T$ is a composite
operation of convolution and down-sampling, color image
demosaicing when $T$ is a color filter array (CFA) masking
operation, and image inpainting when $T$ is the orthogonal projection onto the linear space of matrices. In this paper, we proposed a general model for spherical image inpainting with a new denoiser.

Regarding the degradation equation (\ref{degraded}), the IR task model can be solved through the following optimization,
\begin{equation}\label{optimization}
\hat x=\arg\min_x\|y-T(x)\|+\lambda \Phi(x),
\end{equation}
where the first term is the data fitting with $\Vert\cdot\Vert$ usually chosen to be the Frobenius norm, the second term $\Phi (\cdot)$ is an operator playing the role of regularity, and $\lambda$ is a  positive trade-off parameter. With the aid of the half quadratic splitting (HQS) algorithm, by introducing an auxiliary variable, the optimization problem (\ref{optimization}) can be addressed by iteratively solving the following subproblems
\begin{align}
\mathbf{x}_{k}=&\arg\underset{\mathbf{x}}{\min }\|\mathbf{y}-\mathcal{T}(\mathbf{x})\|^{2}+\alpha\left\|\mathbf{x}-\mathbf{z}_{k-1}\right\|^{2}\label{sub-problem-1}, \\
\mathbf{z}_{k}=&\arg\underset{\mathbf{z}}{\min} ~
 \alpha\left\|\mathbf{z}-\mathbf{x}_{k}\right\|^{2}+\lambda\Phi(\mathbf{z}).
\label{sub-problem-2}
\end{align}
Here $\alpha$ will be set accordingly to specific problems. Equation (\ref{sub-problem-1}) is usually interpreted as the data fitting subproblem and (\ref{sub-problem-2}) as the regularization subproblem.
Many research efforts have been devoted to this hot topic and achieved extensive improvements \cite{chang2021overlapping,dong2019fixing,huang2022quaternion,wong2022incorporating,wu2022efficient}.


In recent years,  deep learning-based models have extensively emerged and achieved 
state-of-the-art restoration performance \cite{cheng2022snow,li2022adjustable,li2022multiple,malgouyres2019multilinear}. The SeaNet proposed in [6] consists of three sub-nets for single image super-resolution with the help of image soft edge. Liu et al. \cite{mwcnn} proposed MWCNN for image restoration, which is a U shape network with DWT and IWT for downsampling and upsampling, respectively, and thus there is no information loss during subsampling.
Both approaches achieved competitive performance in IR tasks.  

To improve interpretability and effectively use the trained neural networks from various tasks, Plug and Play is one of the choices to combine neural networks and prior knowledge of images with an optimization model. Zhang et al. \cite{ ZhangK2021,ircnn2017} developed the deep prior to handling the IR tasks, named plug-and-play (PnP). Specifically, they regarded the regularization term $\Phi (x)$ as a deep denoiser with the deep CNNs. The optimization problem (\ref{optimization}) was solved by the half quadratic splitting (HQS) algorithm and divided into two subproblems, in which the solution of one of the problems is replaced by the deep CNN, which is the so-called deep denoiser. 

Furthermore, the term  (\ref{sub-problem-2}) is usually termed as denoiser prior and conducted by a single CNN denoiser \cite{Ng2020}, which is trained specifically for denoising prior and to replace solving (\ref{sub-problem-2}) to exploit the advantages of CNN.
Following this line, the PnP-based model has wide applications \cite{hou2022truncated}. For example, Wu et al. \cite{Zeng2022} proposed a deep CNN-based PnP framework with MWCNN and has competitive performance in Cauchy noise removal. Zhao et al. \cite{Ng2020} suggested a PnP model for image completion with a low rankness assumption. Fang and Zeng \cite{fang2020} applied the soft edge network \cite{liedge} as a denoiser for image deblurring and denoising and gave a mathematical interpretation of the PnP-based model. Overall, the PnP-based framework has a promising performance.

Many research efforts have been devoted to this hot topic and achieved extensive improvements. Particularly, in recent years,  deep
Plug-and-Play (DPnP) methods have been extensively developed and achieved the
state-of-the-art restoration performance, see for instance \cite{Venkatakrishnan2013, Teodoro2016,Chan2017,Wei2020}. Such a hybrid learning strategy is plugging an off-the-shelf image denoiser
to  resolve IR problems via optimizing the following regularization framework:
\begin{equation}\label{optimization}
\hat x=\arg\min_x\|y-T(x)\|+\lambda \Phi(x),
\end{equation}
where $\Phi$ is an operator playing the role of regularity. 
To implement the optimization, generally two subproblems: the data fitting subproblem and the
regularization subproblem is iteratively solved with the aid of certain optimization
techniques, such as the alternating direction method of multipliers (ADMM) \cite{Admm} and the half quadratic splitting
(HQS) \cite{Hqs}.

%
%
%
%


Motivated by the advantages of the aforementioned PnP models, in this paper, we are going to apply them to image inpainting problem for spherical signals. Precisely, for a spherical signal, with its partially observed samplings,   a novel PnP model integrating spherical framelet decomposition is proposed to  restore the signal. The proposed model is based on low rank assumption under directional spherical Haar tight framelet, which is designed for testing image texture. In addition, we  exploit a newly designed deep convolutional neural network to be the plug-and-play prior denoiser. The network inspired by  \cite{li2022convolutional} and  \cite{jha2020doubleu} employs two-stage encoder-decoder architecture, which is termed as Double-S2HaarNet. Under ground-truth supervision at each stage it provides  progressive and improved denoising.

The rest of this paper is organized as follows. In section \ref{sec:pri}, the related works about spherical signal sampling and frame decomposition are reviewed. The proposed scheme and numerical algorithm are given in section \ref{sec:alg}. Numerical results including gray image and color image inpainting are listed in section \ref{sec:ex}. Section \ref{sec:remarks} concludes this paper.  

\section{Spherical  signal sampling and frame decomposition}\label{sec:pri}

We employ a Haar tight framelet transform that developed in \cite{li2022convolutional}. 
Let $L_2(\mathbb{S}^2)$ be a Hilbert space with inner product $\langle\cdot,\cdot\rangle$ and norm $\|\cdot\|$ defined by
\begin{align*}
	\langle f, g \rangle :=  \int_{\Omega} f(x)g(x) dx ,\\
	\|f\| =  \left ( \int_{\Omega} |f(x)|^2 dx \right)^{\f{1}{2}},
\end{align*}
where $f,g \in L_2(\sph)$ and $\sph \in \RR^3$ is the unit sphere. We call a countable collection $\{e_k\}_{k\in\Lambda}\subset L_2(\sph)$  a \emph{tight frame} with frame bound $c$ if there exists a constant $c>0$ such that
\begin{align*}
f = \f{1}{c} \sum_{k\in\Lambda}  \langle f, e_k \rangle e_k \quad \forall \ f\in  L_2(\sph).
\end{align*}
The frame decomposition is a transformation $\h$ given by 
\begin{align*}
	\h: f\in L_2(\sph) \rightarrow \{ \langle f, e_i \rangle: e_i \in \{e_k\}_{k\in\Lambda} \},
\end{align*}
and the reconstruction $\h^*$
\begin{align*}
\h^*:  \{ \langle f, e_i \rangle: e_i \in \{e_k\}_{k\in\Lambda} \} \rightarrow f\in L_2(\sph).
\end{align*}


 A Haar tight frame on the sphere can be constructed based on a hierarchical partition.

\begin{definition}
	Let $ \NN_0$ be a set of nonnegative integers. We call  $\{\mathcal{B}_j\}_{j\in \NN_0}$
	a \emph{hierarchical partition} of $\sph$ if the following three conditions are satisfied:
	\begin{enumerate}
		\item[{a)}] Root property: $\mathcal{B}_0 = \{\sph \}$ and each $\mathcal{B}_j$
		is a partition of $\sph$ having finitely many measurable sets with positive measures.
		\item[{b)}] Nested property: for any $j\in \NN$ and any (child) set $R_1 \in \mathcal{B}_j$, there exists a (parent) set $R_0 \in \mathcal{B}_{j-1}$ such that $R_1 \subseteq R_0$. In other word, partition $\mathcal{B}_j$ is a refinement of the partition $\mathcal{B}_{j-1}$.
		\item[{c)}] Density property: the maximal diameters among the sets in $\mathcal{B}_j$ tend to zero as $j$ tends to infinity.
	\end{enumerate}
\end{definition}

Denote  $\Lambda_j:= [\ell_1]\times\cdots\times[\ell_j]$ to be an index set for the labeling sets in $\CB_j$, where $[N] = \{1,\dots,N\}$ for any positive integer $N$ and
\[
\CB_j=\{R_{\vec v}\subseteq \mathbb{S}^2, \,\,\, \vec v\in \Lambda_j\}.
\]

By the nested property,  $ R_{(\vec v,i)}\subseteq R_{\vec v}$ for $\vec v\in\Lambda_{j-1}$ and $i\in[\ell_j]$.  Now, for each  $\vec v\in \Lambda_j$, define a Haar-type scaling function to be
\begin{equation}
\label{eq:phi}
	\phi_{\vec v}:=\frac{\chi_{R_{\vec v}}}{\sqrt{|R_{\vec v}|}},
\end{equation}
and for some integer $n_j\geq 1$,   $n_j$ Haar-type framelet functions to be
\begin{equation}\label{eq:Psi}
	\psi_{(\vec v,k)}=\sum_{i\in[\ell_j]}{a^{(\vec v)}_{k,i}}\phi_{(\vec v,i)},\quad k=1,\ldots, n_j,
\end{equation}
where $a^{(\vec v)}_{k,i}$ is the $(k,i)$-entry  of some matrix $\b A_{\vec v}\in\RR^{n_j\times l_j}$.
By setting proper matrices $A_{\vec v}$, one can construct a Haar-type tight frame and develop its fast decomposition and reconstruction algorithms. The following corollary determines the framelet and algorithm we shall use.

\begin{corollary}\label{cor}
    There exists a collection $\{\phi_{\vec u}\}_{\vec u\in \Lambda_L} \cup \{\psi_{(\vec v,k)}, k\in[6]\}_{j \geq L, \vec v \in \Lambda_j}\subset L_2(\sph)$ determined by a hierarchical partition with each parent containing four children that forms a Haar tight frame with frame bound $1$, and the corresponding operators $\h$ and $\h^*$ depend on the following matrix $P$
	\[
	P =\f 1 { 2}
	\begin{bmatrix}
	1 & 1 & 1 & 1 \\
	1 & -1 & 0 & 0 \\
	1 & 0 & -1 & 0 \\
	1 & 0 & 0 & -1 \\
	0 & 1 & -1 & 0 \\
	0 & 1 & 0 & -1 \\
	0 & 0 & 1 & -1
	\end{bmatrix}.
	\]
\end{corollary}

\begin{figure}[htp]
	\centering
	\includegraphics[width=3in]{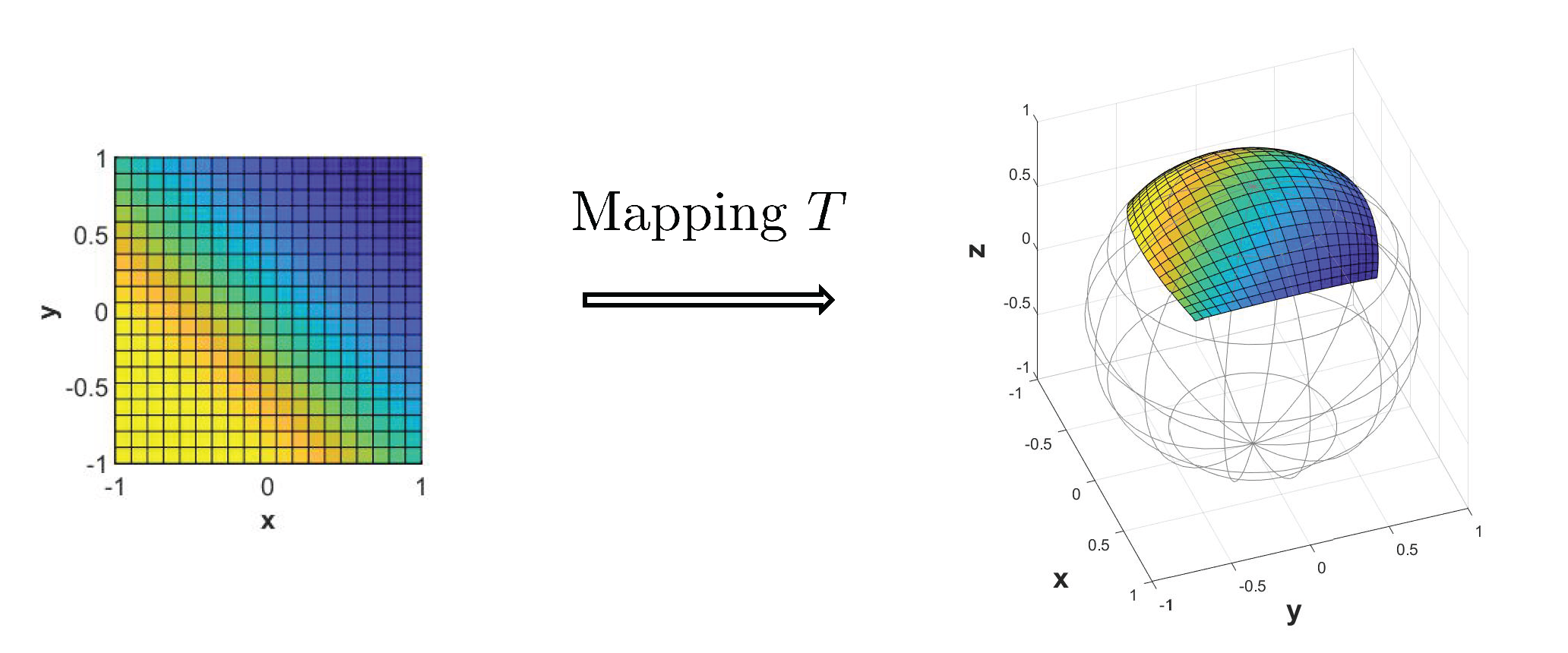}
	\caption{Visualization of mapping $T$
which maps a square to a spherical cap.}
	\label{fig:T}
\end{figure}

As it is well known, computers can only deal with discrete signals. To do spherical signal processing, we first need a proper way to discretize an analog signal. In this work, we take the discretization sampling method based on an area-regular partition of 2-sphere \cite{li2022convolutional}.
It was constructed through a bijective mapping and its rotations:
$ T:[-1, 1]\times[-1, 1] \rightarrow  \sph $ defined by $T(x,y) = \frac{(x, y,1)}{\sqrt{x^2+y^2+1}}$.
See Figure \ref{fig:T} below for the illustration.

\begin{figure}[htp]
	\centering
	\includegraphics[width=4in]{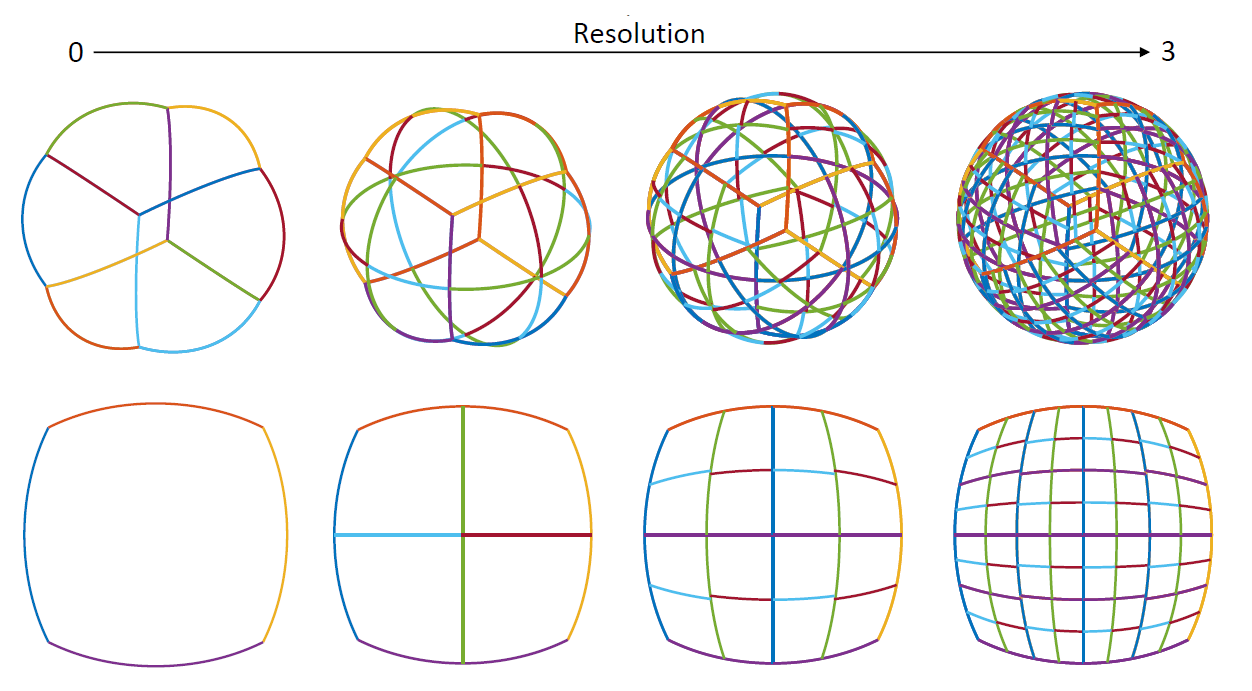}
	\caption{Partition Process.}
	\label{Partitions}
\end{figure}

Then for any given resolution $J\geq 0$, a $2$-sphere can be divided into equal-area partitions, see Figure \ref{Partitions} for illustration. This forms an algorithm for a hierarchical partition on the 2-sphere.

By taking the centers of the partition patches, the samplings of an analog signal can be distributed equivalently, which takes advantage over the traditional spherical coordinates discretization.

Based on the above discussion, any signal $f \in L_2(\sph)$ is discretized to $\b f$, which depends on a certain resolution $J$.  The discrete signal $\b f$ is actually the set $\{f(\b x_i): \b x_i \in S_i, S_i \in \mathcal{B}_j, \bigcup_{i}S_i =  \sph, S_j\cap S_k = \empty, \forall j \neq k  \}$. We assume that the dataset is defined on some resolution level in the following. Applying the discretization and \corref{cor}, the spherical Haar framelet and fast framelet transform algorithm are exactly constructed.

With the help of the fast decomposition and reconstruction algorithms, it, on the one hand, allows our model to capture directional texture details. On the other hand, it can reduce the spatial footprint and granularity of convolutions.

\begin{figure}[htp]
	\centering
	\includegraphics[width=5in]{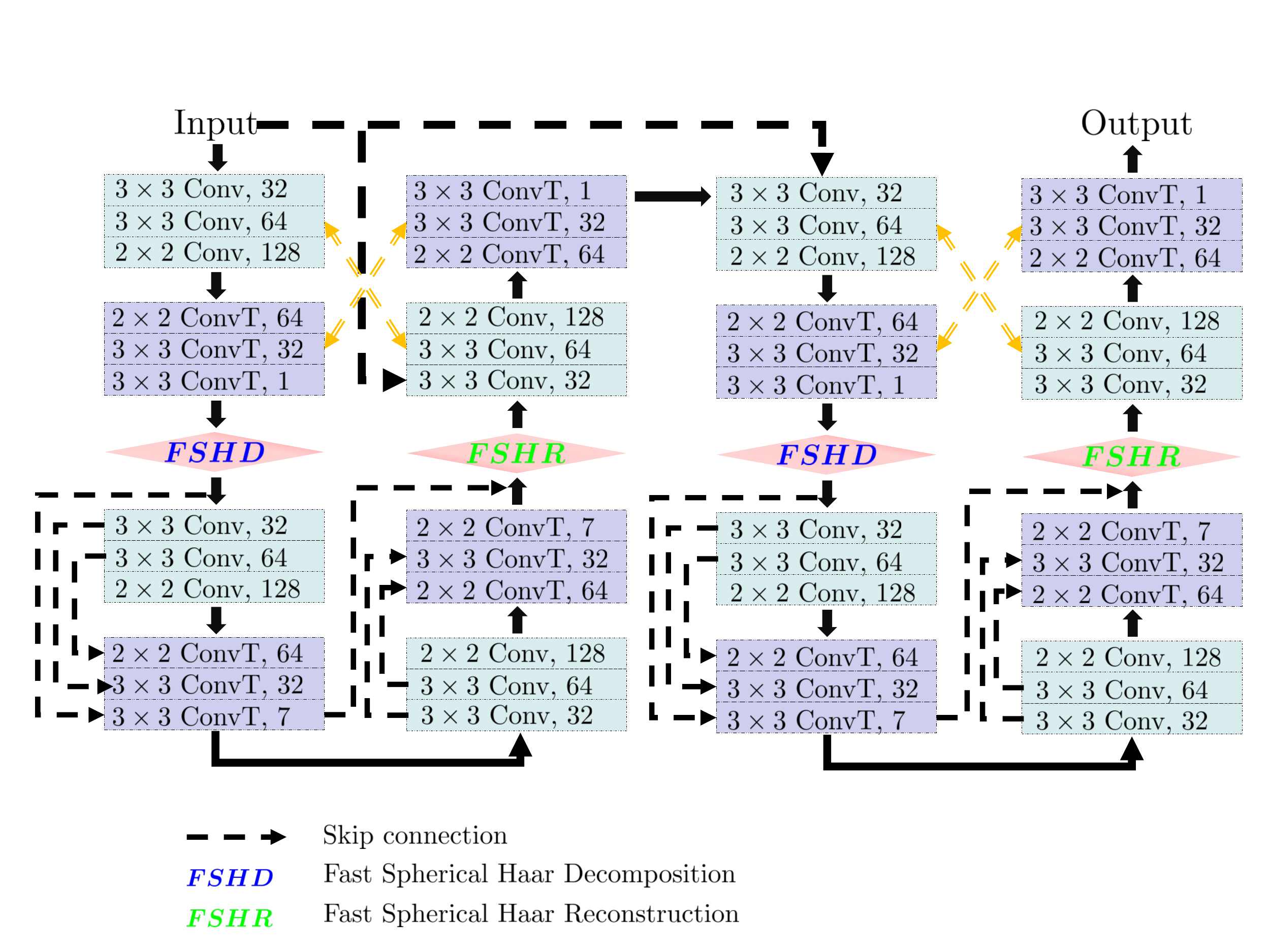}
	\caption{Double-S2HaarNet.}
	\label{doubleHaar}
\end{figure}

\section{The proposed model and algorithm}
\label{sec:alg}

In this section, based on the aforementioned
off-the-shelf spherical signal sampling and Haar-type framelets, to enhance the inpainting performance, we shall first improve the denoiser and then exploit a plug-and-play model involving fast frame decomposition.

\subsection{Improved Denoiser}\label{3.1}

As mentioned above, the iteration \ref{sub-problem-2} will be conducted by a denoiser. The performance of the denoiser will affect the resulting restoration.
In \cite{li2022convolutional}, a CNN spherical denoising model, S2HaarNet, was developed and achieved a competitive performance.
In the present paper, we further exploit a new spherical CNN (illustrated in Figure \ref{doubleHaar}), which partly follows the infrastructure of the Double-Unet \cite{jha2020doubleu} and S2HaarNet \cite{li2022convolutional}, and incorporates the skip connections and spherical frame transformations. Thus we shall call it Double-S2HaarNet. The new network consists of two feature encoder-decoder stages for which each one follows S2HaarNet. We then take ground-truth supervision at each stage for progressive and improved denoising performance.
We adopt the feature
concatenation by combining the feature maps
from the encoder path and decoder path,
which can capture multi-scale information and enrich
feature representation for a better feature
prediction.
To bridge the two blocks, we concentrate the input and output of the first block and feed it into the second. Our model can also be readily extended to deal with color images by handling three channels independently.


\subsection{Proposed PnP model}
To develop a PnP model, besides the above pre-trained denoiser,  a proper design for data fitting subproblem plays a crucial role as well.  In this work,  we attempt to propose a new data fitting operator to simultaneously exploit the strengths of the PnP model.  In \cite{Ng2020}, data fitting was operated by using the tensor singular value decomposition and tensor nuclear norm, which promoted the low-rankness of the underlying tensor. Motivated by such an idea, we instead utilize the tight frame decomposition in our data fitting and suggest the novel PnP model as follows:
\begin{equation}\label{w1}
\begin{aligned}
&\min_{\b x} \|\h \b x\|_0 + \lambda \Phi(\b x)\\
&\text{s.t.~~} \
 P_{\sph}(\b x) = P_{\sph}(\b g),
\end{aligned}
\end{equation}
where  $\h x$ is the coefficients of tight frame decomposition as mentioned in section \ref{3.1}, $ \Phi(\b x)$ is an implicit regularizer by plugging our denoising Double-S2HaarNet, and $\lambda$ is a positive parameter.

\subsection{Implementation Details}


We apply ADMM framework to solve the optimization problem. Notice that in practice, it is more convenient to replace $\|\cdot\|_0$ by $\|\cdot \|_1$.
First, we denote the indicator function as
\begin{equation}
\mathbf{1}_{\mathbb{S}}(\b x)=\left\{\begin{array}{ll}
0, & \text { if } \b x\in\{\b x~\vert~{P}_{\sph}(\b x)={P}_{\sph}({\b g})\}. \\
\infty, & \text { otherwise }.
\end{array}\right.
\end{equation}
Then we reformulate model (\ref{w1}) as 
\begin{equation}\label{w2}
\begin{aligned}
&\min _{\b x}\| \b y\|_{1}+\lambda \Phi(\b z)+\mathbf{1}_{\mathbb{S}}(\b x) \\
&\text { s.t. } \quad \b y=\h \b x, \b z=\b x.
\end{aligned}
\end{equation}
The augmented Lagrangian function of (\ref{w2}) is 
\begin{equation}\label{lag}
\begin{aligned}
\y{\mathcal{L}(\b x, \b y, \b z; \Lambda_1, \Lambda_2)=}
\|\b y\|_{1}+\lambda \Phi(\b z)+\mathbf{1}_{\mathbb{S}}(\b x)&+\left\langle \b y-\h \b x, \Lambda_{1}\right\rangle+\frac{\beta_{1}}{2}\|\b y-\h \b x\|^{2}
+\left\langle \b z- \b x, \Lambda_{2}\right\rangle+\frac{\beta_{2}}{2}\|\b z-\b x\|^2,
\end{aligned}
\end{equation}
\y{
where $\beta_1, \beta_2>0$ are two penalty parameters and $\Lambda_1, \Lambda_2$ are the Lagrange multipliers. The ADMM iteration for solving (\ref{lag}) goes as follows,
\begin{equation}
\left\{\begin{array}{l}
y=\arg \min _{\b y}\|\b y\|_{1}+\left\langle \b y-\h \b x, \Lambda_{1}\right\rangle+\frac{\beta_{1}}{2}\left\|\b y-\h \b x\right\|^{2}, \\
\b z=\arg \min _{\b z} \lambda \Phi(\b z)+\left\langle \b z-\b x, \Lambda_{2}\right\rangle+\frac{\beta_{2}}{2}\|\b z-\b x\|^{2}, \\
\b x=\arg \min _{\b x} \mathbf{1}_{\mathbb{S}}(\b x)+\frac{\beta_{1}}{2}\Vert\b y-\h \b x+\frac{\Lambda_{1}}{\beta_{1}}\Vert^{2}+\left\langle \b z-\b x, \Lambda_{2}\right\rangle+\frac{\beta_{2}}{2}\|\b z-\b x\|^{2},\\
\Lambda_1=\Lambda_1+(\b y-\h \b x),\\
\Lambda_2=\Lambda_2+(\b z-\b x).
\end{array}\right.
\end{equation}
Next, we elaborate on how to solve these subproblems respectively.
}

\begin{itemize}
	\item \y{The $\b y$-subproblem is written as}
	\begin{equation}
	\begin{aligned}
	\b y &=\arg \min _{\b y}\|\b y\|_{1}+\left\langle \b y-\h \b x, \Lambda_{1}\right\rangle+\frac{\beta_{1}}{2}\left\|\b y-\h \b x\right\|^{2} \\
	&=\arg \min _{\b y}\|\b y\|_{1}+\frac{\beta_{1}}{2}\left\|\b y-\h \b x+\frac{\Lambda_{1}}{\beta_{1}}\right\|^{2}.
	\end{aligned}
	\end{equation}
	\y{Then the solution of $y$ can be obtained by}
	\begin{equation}
	    \begin{aligned}
	\b y &=\operatorname{shrink}\left(\h \b x-\frac{\Lambda_{1}}{\beta_{1}}, \frac{1}{\beta_{1}}\right) \\
	&=\max \left(\left\|\h \b x-\frac{\Lambda_{1}}{\beta_{1}}\right\|_{2}-\frac{1}{\beta_{1}}, 0\right) \frac{\h \b x-\frac{\Lambda_{1}}{\beta_{1}}}{\left\|\h \b x-\frac{\Lambda_{1}}{\beta_{1}}\right\|},
	\end{aligned}
	\end{equation}
	\y{where the $\operatorname{shrink}$ operator is a soft shrinkage operator.}
	\item \y{The $\b z$-subproblem is written as}
	\begin{equation}\label{denoiser}
	\begin{aligned}
	\b z &=\arg \min _{\b z} \lambda \Phi(\b z)+\left\langle \b z-\b x, \Lambda_{2}\right\rangle+\frac{\beta_{2}}{2}\|\b z-\b x\|^{2} \\
	&=\arg \min _{\b z} \lambda \Phi(\b z)+\frac{\beta_{2}}{2}\left\|\b z-\b x+\frac{\Lambda_{2}}{\beta_{2}}\right\|^{2},
	\end{aligned}
	\end{equation}
	{According to Bayes rule, Eq. (\ref{denoiser}) corresponds to denoising the image $\b x-\Lambda_2/\beta_2$ by the CNN denoiser with noise level $\lambda/\beta_2$. To address this, we rewrite Eq. (\ref{denoiser}) as
	\begin{equation}
	   \b z=\mbox{Denoiser}(\b x-\frac{\Lambda_2}{\beta_2},\sqrt{\frac{\lambda}{\beta_2}}).
	\end{equation}
	In this paper, we apply the Double-S2HaarNet as the denoiser.
	}
	\item \y{The $\b x$-subproblem is written as}
	\begin{equation}\label{x1}
	\begin{aligned}
	\b x &=\arg \min _{\b x} \mathbf{1}_{\mathbb{S}}(\b x)+\frac{\beta_{1}}{2}\left\|\b y-\h \b x+\frac{\Lambda_{1}}{\beta_{1}}\right\|^{2}+\left\langle \b z-\b x, \Lambda_{2}\right\rangle+\frac{\beta_{2}}{2}\|\b z-\b x\|^{2} \\
	&=\arg \min _{\b x} \mathbf{1}_{\mathbb{S}}(\b x)+\frac{\beta_{1}}{2}\left\|\b y-\h \b x+\frac{\Lambda_{1}}{\beta_{1}}\right\|^{2}+\frac{\beta_{2}}{2}\left\|\b z-\b x+\frac{\Lambda_{2}}{\beta_{2}}\right\|^{2}.
	\end{aligned}
	\end{equation}
	\y{By minimizing the $\b x$-subproblem, we have $\mathbf{1}_{\mathbb{S}}(\b x)=0$, i.e., $\b x\in\mathbb{S}$. Then optimality condition of (\ref{x1}) is given by
	\begin{equation}
	    \beta_1\h^*(\h\b x-\b y-\frac{\Lambda_1}{\beta_1})+\beta_2(\b x-\b z-\frac{\Lambda_2}{\beta_2})=0.
	\end{equation}
	Since $\h^*\h=I$,we obtain the following linear system,
	\begin{equation}
	    (\beta_{1}+\beta_{2})\b x =\beta_{1} \h^{*} \b y+\beta_{2} \b z+\h^{*} \Lambda_{1}+\Lambda_{2}.
	\end{equation}
	Thus, the closed-form solution of $\b x$-subproblem is given as follows:
\begin{equation}
\left\{\begin{array}{l}
\mathcal{P}_{\sph}\left(\b x\right)=\mathcal{P}_{\sph}(\b g), \\
\mathcal{P}_{(\sph)^{c}}\left(\b x\right)=\mathcal{P}_{(\sph)^{c}}\left(\frac{\beta_{1} \h^{*} \b y+\beta_{2} \b z+\h^{*} \Lambda_{1}+\Lambda_{2}}{\beta_{1}+\beta_{2}}\right),
\end{array}\right.
\end{equation}
where $(\sph)^{c}$ denotes the complementary set of $\sph$.}
\end{itemize}

\begin{figure}[h]
	\centering
	\subfigure[Barbara]{\includegraphics[width=2.5cm]{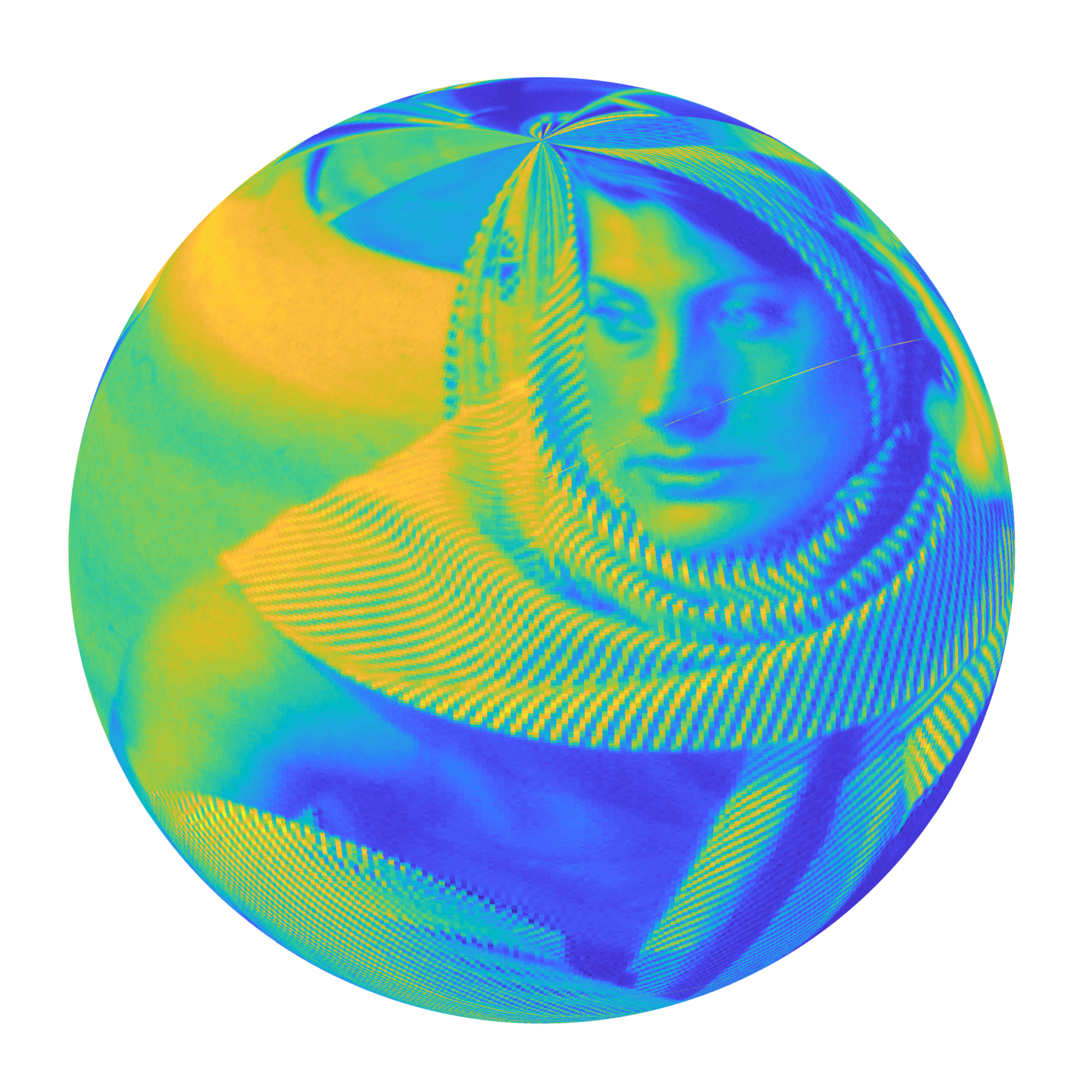}}
	\subfigure[Boat]{\includegraphics[width=2.5cm]{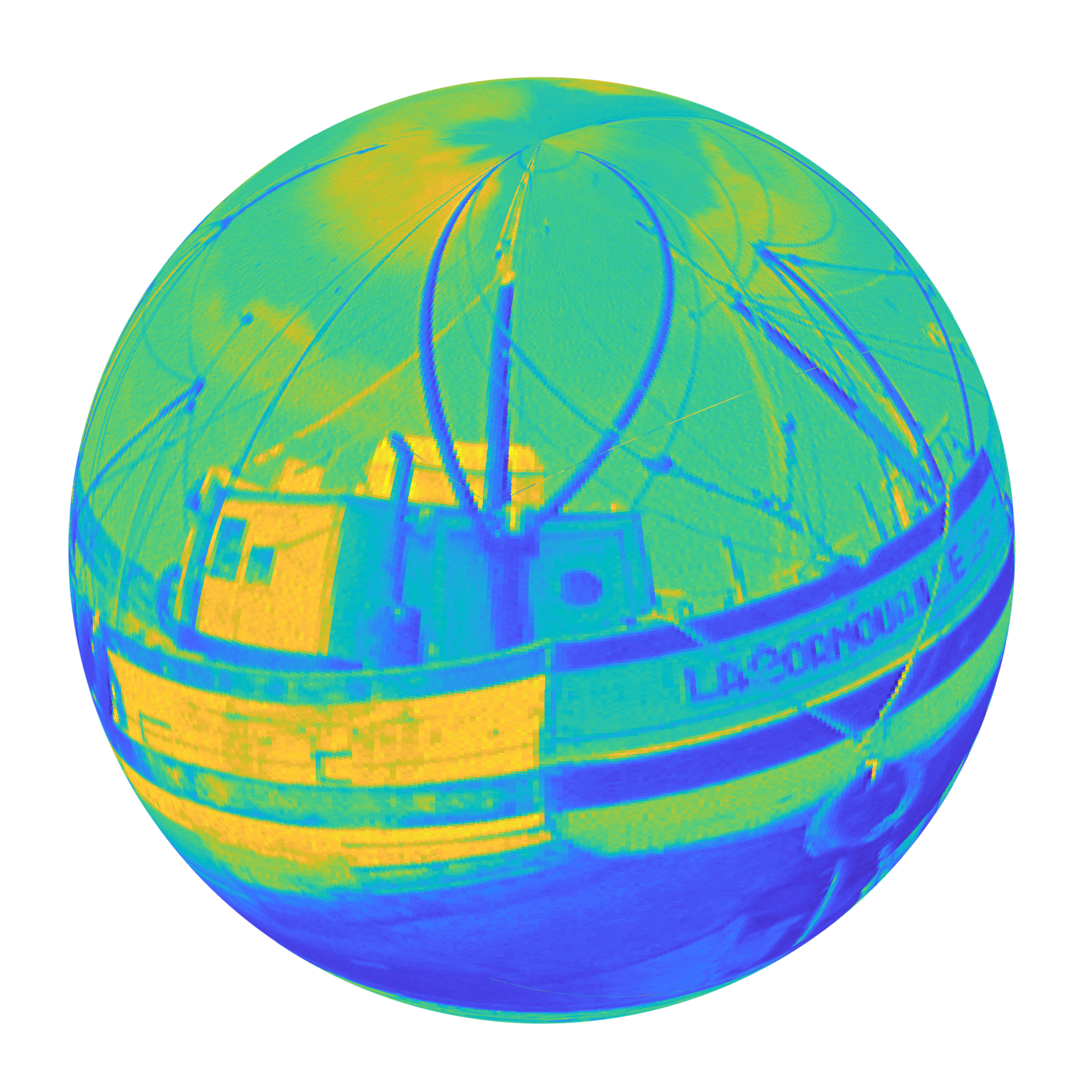}} \\
	\subfigure[Fingerprint]{\includegraphics[width=2.5cm]{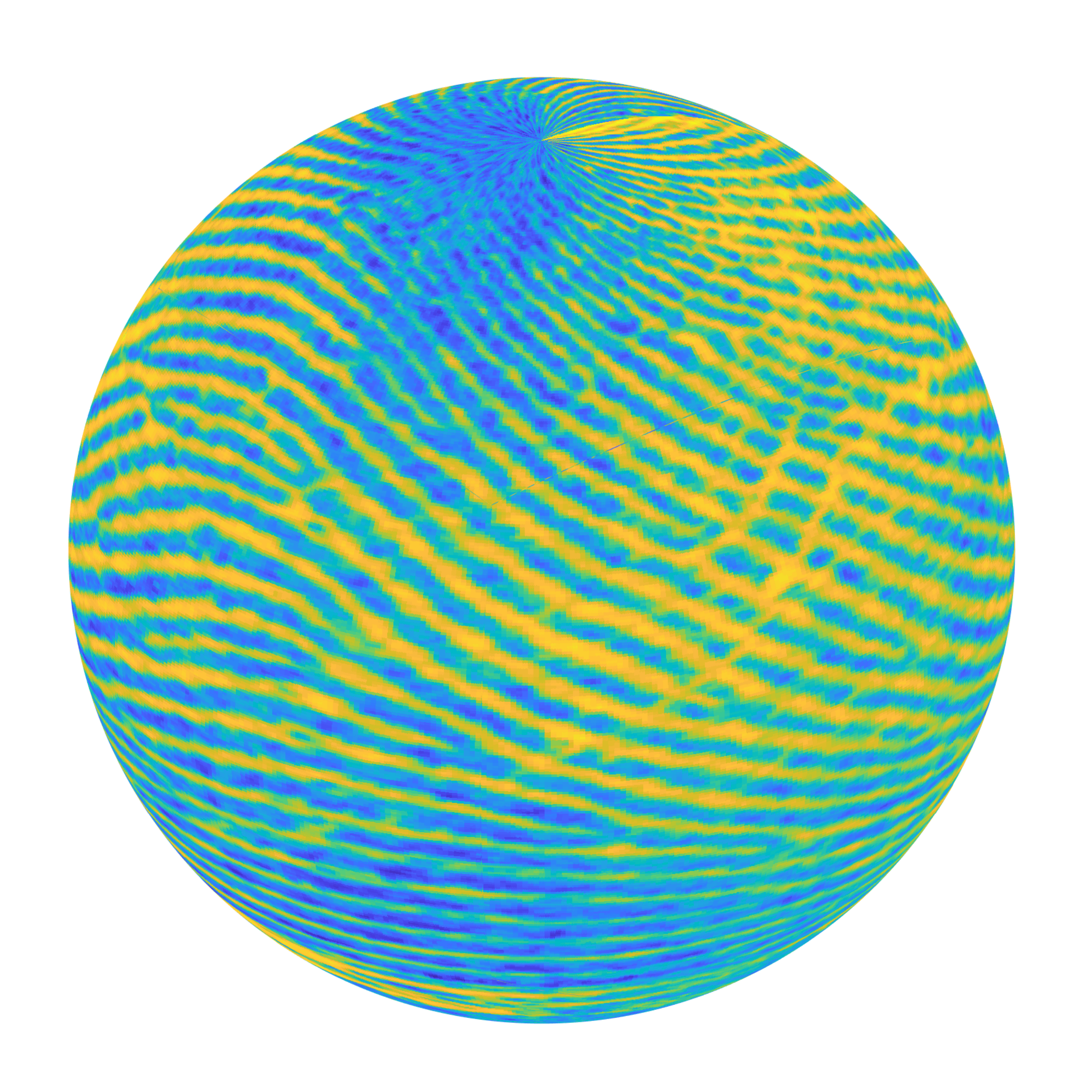}}
	\subfigure[Hill]{\includegraphics[width=2.5cm]{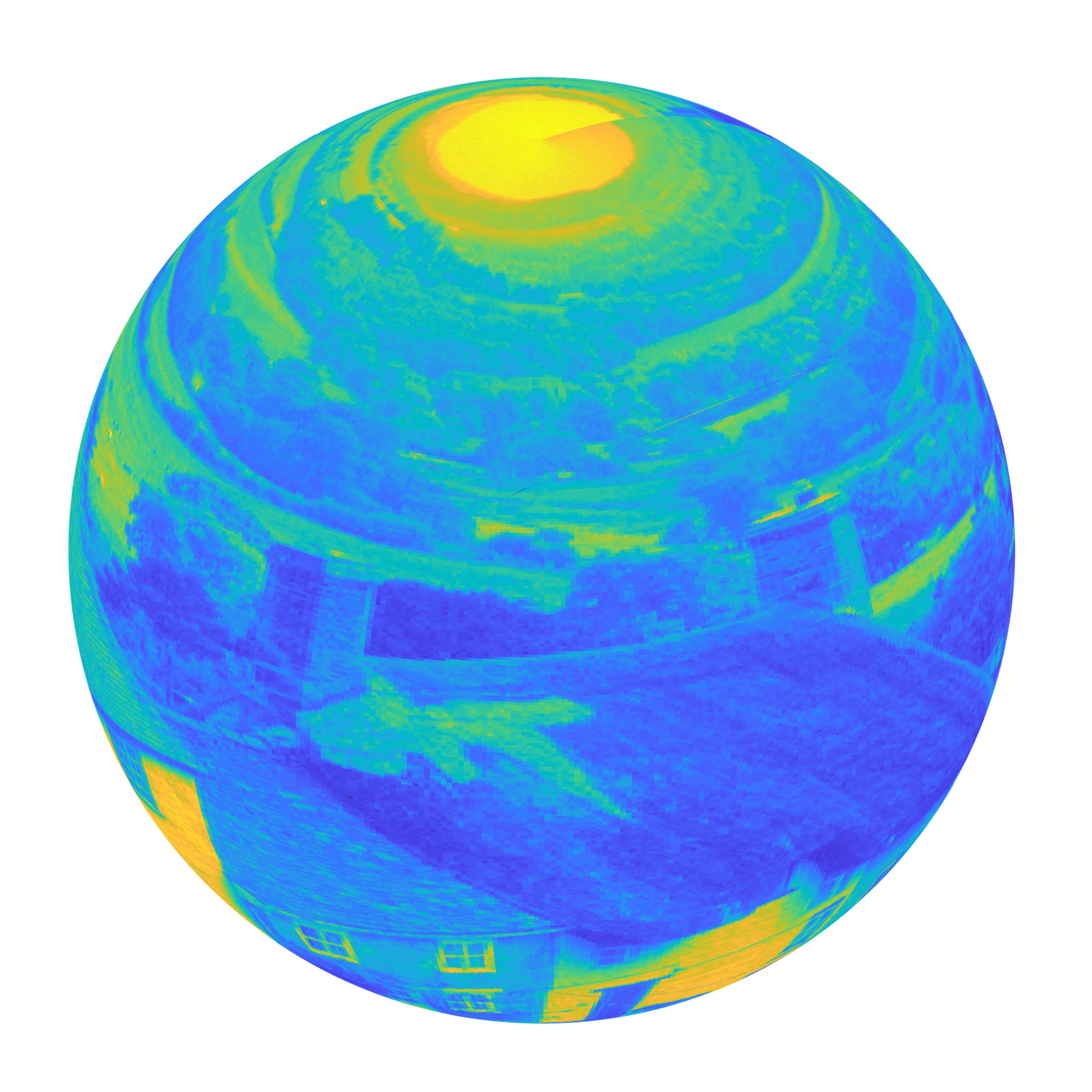}}
	\subfigure[Man]{\includegraphics[width=2.5cm]{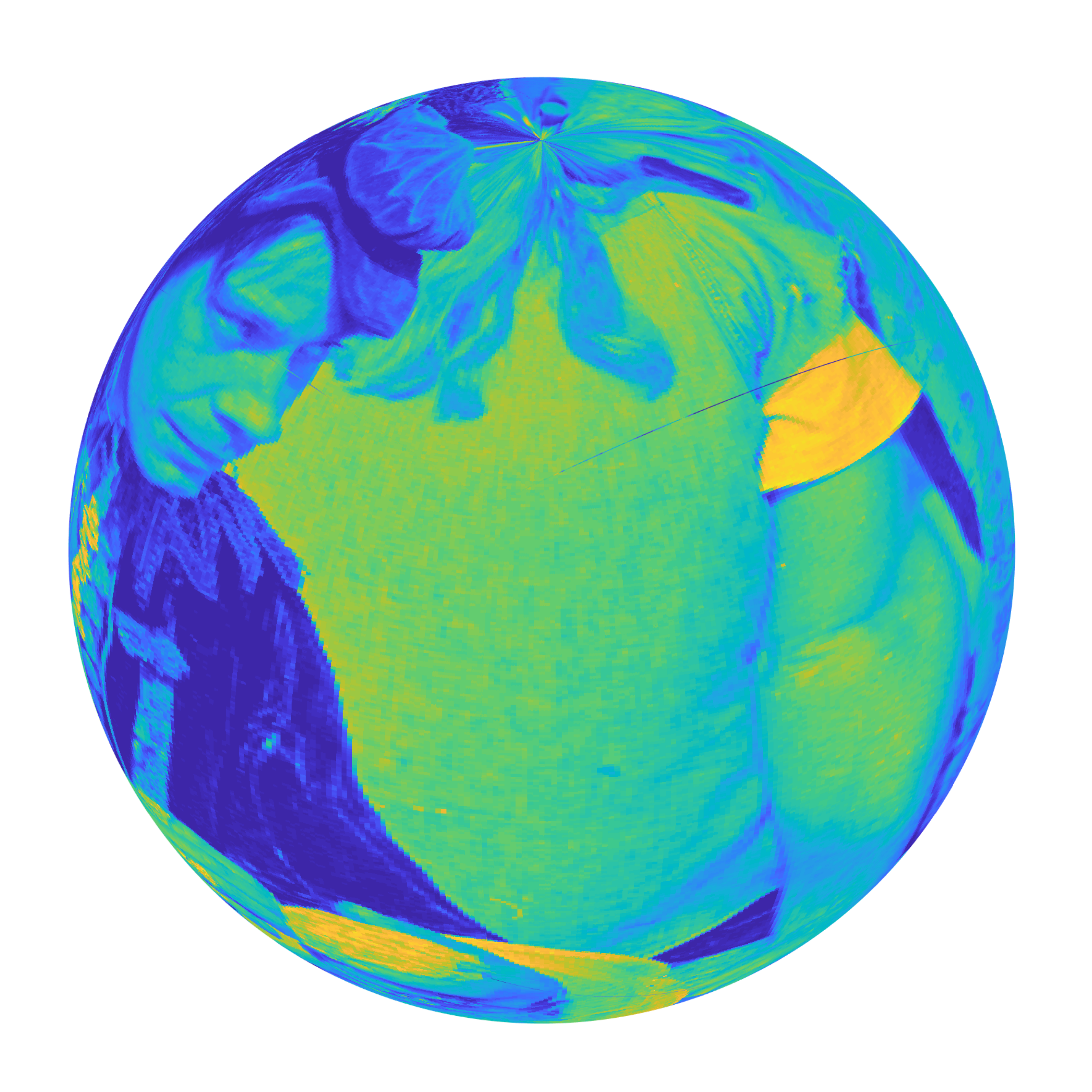}}
	\caption{Five grey images for testing.}
\end{figure}


\begin{table}[h]
	\centering
	\footnotesize
	\setlength\tabcolsep{1.5pt}
	\linespread{1.2}
	
	\setlength{\tabcolsep}{10pt} 
	\renewcommand{\arraystretch}{1.5} 
	\caption{{Average inpainting results with PSNR/SSIM on F-360iSOD. In Fx, x = 6,7,8, represents the resolution level. The best results are highlighted.}}\label{t1}
	\begin{tabular}{c|c|c|c|c|c}
		\hline 
		dataset&methods\ &  50\% &80\% & 90\%  &95\%  \\ 
		\hline
		\multirow{5}{*}{F6}&degraded&10.08/0.2087&8.04/0.0806&7.52/0.0445&7.29/0.0268\\
		\cline{2-6}&S2HaarNet&23.02/0.7711&20.27/0.5777&19.16/0.4962&18.02/0.4175\\
		\cline{2-6}&S2HaarNetPnP&23.96/0.8125&20.81/0.6024&19.47/0.5225&18.33/0.4290\\
		\cline{2-6}&DoubleS2HaarNet&24.22/0.8321&20.99/0.6470&19.55/0.5367&18.41/0.4568\\
		\cline{2-6}&DoubleS2HaarNetPnP&\bf{24.63/0.8470}&\bf{21.49/0.6756}&\bf{19.96/0.5495}&\bf{18.63/0.4596}\\
		\hline
		\multirow{5}{*}{F7}&degraded& 10.08/0.1857 &8.04/0.0776 &7.53/0.0454 &7.29/0.0290  \\ 
		\cline{2-6}&S2HaarNet&23.94/0.7822&21.12/0.5920&20.03/0.5125&18.89/0.4380\\
		\cline{2-6}&S2HaarNetPnP&24.92/0.8213&21.72/0.6200&20.33/0.5445&19.01/0.4553\\
		\cline{2-6}
		&DoubleS2HaarNet&25.15/0.8349&21.83/0.6590&20.41/0.5506&19.30/0.4750\\
		\cline{2-6}
		&DoubleS2HaarNetPnP&\bf{26.14/0.8623} &\bf{22.63/0.7009} &\bf{20.93/0.5901}&\bf{19.60/0.5007}\\ 
		\hline 
		\multirow{5}{*}{F8}&degraded& 10.08/0.1660&8.04/0.0755& 7.53/0.0469 &7.29/0.0313\\ 
		\cline{2-6}&S2HaarNet&26.86/0.8402&23.79/0.6965&22.49/0.6337&21.05/0.5633\\
		\cline{2-6}&S2HaarNetPnP&27.35/0.8613&24.18/0.7313&22.84/0.6667&21.49/0.5953\\
		\cline{2-6}
		&DoubleS2HaarNet&28.55/0.8918&24.76/0.7607&23.03/0.6751&21.64/0.6006\\
		\cline{2-6}
		&DoubleS2HaarNetPnP&\bf{28.93/0.8961}&\bf{25.02/0.7713}&\bf{23.31/0.6862}&\bf{21.87/0.6167}\\
		\hline
	\end{tabular} 
\end{table}

\section{Experimental results}
\label{sec:ex}
In this section, we present experimental results to verify the performance of the proposed model Double-S2HaarNetPnP in image inpainting. As aforementioned, we proposed a plug-and-play model for the image inpainting task. The parameter of the optimization function (\ref{lag}) are set as $\lambda=1$, $\beta_1 \in [0.1, 1]$ with step $0.1$, $\beta_2 \in [1, 5]$ with step $1$. For training Double-S2HaarNet, we use the ADAM algorithm and a mini-batch size of $16$. The learning rate decays exponentially from the beginning value $0.001$ with a multiplicative factor $0.9$ in $100$ epochs. Weight decay is chosen to be $0.001$. Since the contrast of grayscale images is relatively low, we present the visual effects of the image with color so that the image information can be displayed more clearly.

\subsection{Datasets}
The dataset for training CNN denoisers is produced by applying spherical sampling operation (defined in section \ref{sec:pri}) on the  dataset Caltech101 \cite{fei2004learning} with 7677 for training and 1000 for validation. For the testing, we choose the dataset F-360iSOD \cite{zhang2020fixation} which contains 107 omnidirectional images. Additionally,  we take five classical images as illustrated in Figure 4 for testing as well.


\begin{figure}[h]
	\begin{minipage}{0.3\linewidth}
		\centering
		\centerline{\includegraphics[width=1.6in]{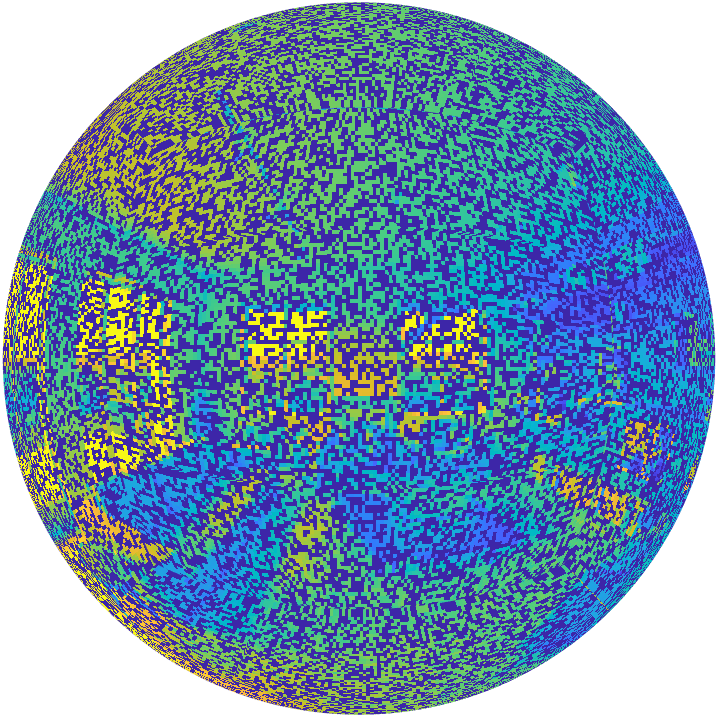}
		}
		\centerline{\small{(a) degraded}}
		\centerline{\small{7.12/0.0532}}
	\end{minipage}
	\begin{minipage}{0.3\linewidth}
		\centering
		\centerline{\includegraphics[width=1.6in]{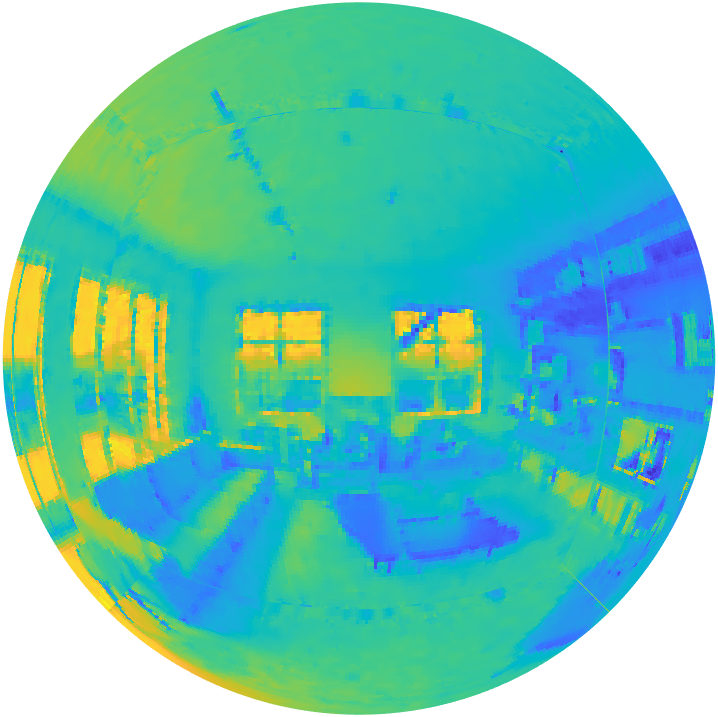}}
		\centerline{\small{(b) HaarNet}}
		\centerline{\small{28.32/0.8735}}
	\end{minipage}
	\begin{minipage}{0.3\linewidth}
		\centering
		\centerline{\includegraphics[width=1.6in]{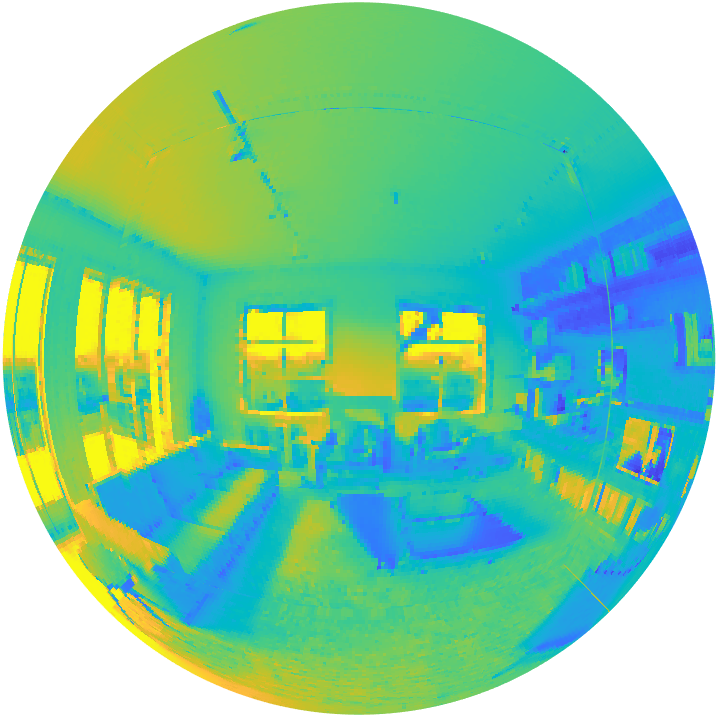}}
		\centerline{\small{(c) HaarNetPnP}}%
		\centerline{\small{30.15/0.9349}}
	\end{minipage}
\vspace{0.1in}
	
	\begin{minipage}{0.3\linewidth}
		\centering
		\centerline{\includegraphics[width=1.6in]{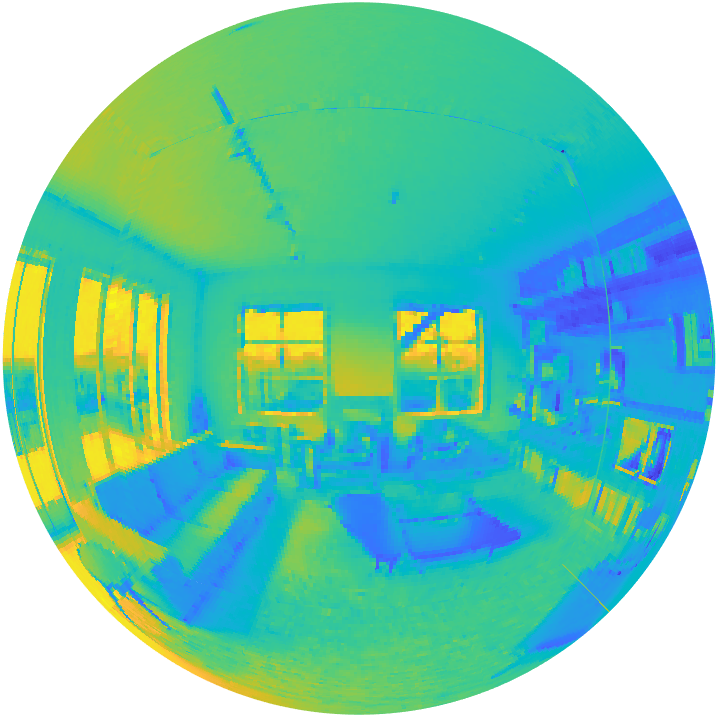}}
		\centerline{\small{(d) DoubleHaarNet}}
		\centerline{\small{30.87/0.9346}}
	\end{minipage}
	\begin{minipage}{0.3\linewidth}
		\centering
		\centerline{\includegraphics[width=1.6in]{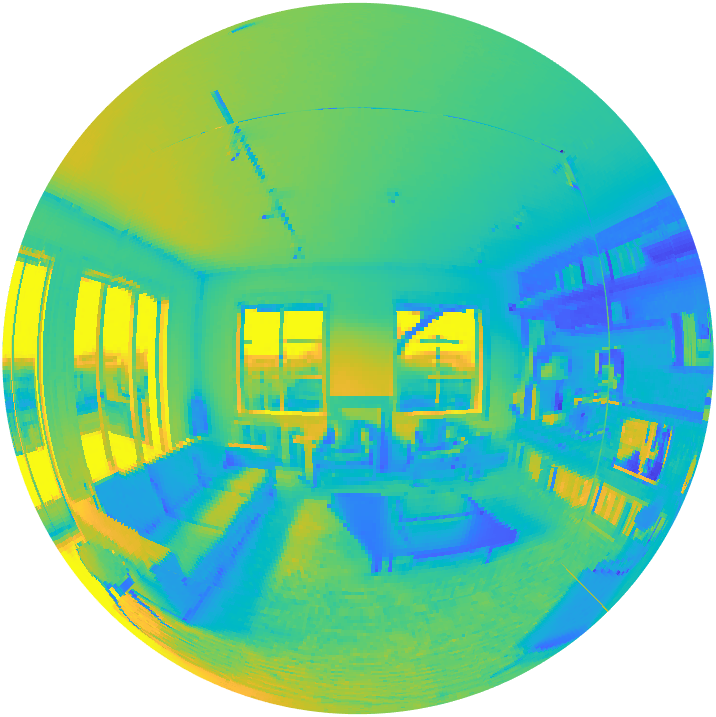}}
		\centerline{\small{(e) DoubleHaarNetPnP}}
		\centerline{\small{31.87/0.95542}}
	\end{minipage}
	\begin{minipage}{0.3\linewidth}
		\centering
		\centerline{\includegraphics[width=1.6in]{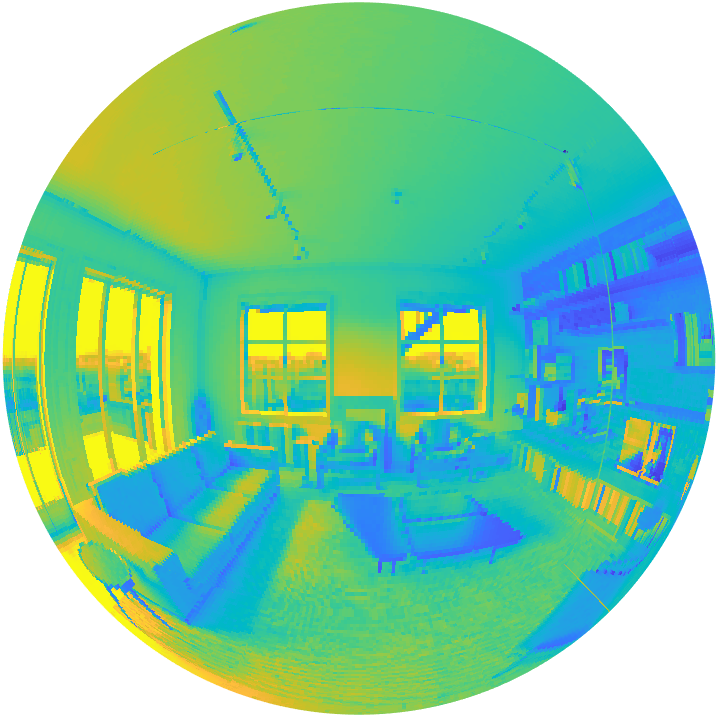}}
		\centerline{\small{(f) Ground-truth}}
		\centerline{\small{infinity/1}}
	\end{minipage}
	\caption{The inpainting results (PSNR (dB)/SSIM) with random missing ratio 50\%. (a) the degraded image; the recovered results of (b) Haar network only; (c) plug and play with Haar network; (d) DoubleHaar network only; (e) plug and play with DoubleHaar network; (f) the original image.}\label{f4}
\end{figure}

\begin{table}{h}
    \centering
    \footnotesize
    \setlength\tabcolsep{1.5pt}
    \linespread{1.2}
    
    \setlength{\tabcolsep}{10pt} 
    \renewcommand{\arraystretch}{1.5} 
    
	\caption{\y{Inpainting results with PSNR/SSIM. The best results are highlighted.}}\label{t2}
	\begin{tabular}{c|c|c|c|c|c}
		\hline 
		Images&methods\ &  50\% &80\% & 90\%  &95\%  \\ 
		\hline
		\multirow{5}{*}{Barbara}&degraded&8.54/0.1039&6.50/0.0466&5.99/0.0281&5.75/0.0169\\
		\cline{2-6}&HaarNet&26.91/0.8523&23.77/0.7092&22.57/0.6386&21.30/0.5653\\
		\cline{2-6}&HaarNetPnP&27.56/0.9043&24.36/0.7547&23.14/0.6834&21.70/0.5731\\
		\cline{2-6}&DoubleHaarNet&28.81/0.9153&24.79/0.7732&23.17/0.6788&21.87/0.6083\\
		\cline{2-6}&DoubleHaarNetPnP&\bf{29.14/0.9283}&\bf{25.15/0.7895}&\bf{23.75/0.6836}&\bf{22.21/0.6799}\\
		\hline
		\multirow{5}{*}{Boat}&degraded&8.57/0.0976&6.52/0.0453&6.02/0.0281&5.78/0.0172\\
		\cline{2-6}&HaarNet&30.33/0.8569&26.94/0.7618&25.08/0.6995&22.84/0.6158\\
		\cline{2-6}&HaarNetPnP&32.04/0.9202&28.08/0.8079&25.77/0.7141&23.16/0.6372\\
		\cline{2-6}&DoubleHaarNet&32.69/0.9208&28.20/0.8099&25.83/0.7371&23.60/0.6554\\
		\cline{2-6}&DoubleHaarNetPnP&3\bf{3.12/0.9299}&\bf{28.81/0.8234}&\bf{26.63/0.7567}&\bf{23.84/0.6946}\\
		\hline
		\multirow{5}{*}{Fingerprint}&degraded&7.39/0.1268&5.35/0.0477&4.83/0.0254&4.60/0.0136\\
		\cline{2-6}&HaarNet&27.53/0.9165&23.88/0.8183&21.28/0.7101&17.90/0.4914\\
		\cline{2-6}&HaarNetPnP&29.10/0.9465&24.62/0.8294&21.79/0.7342&18.39/0.5394\\
		\cline{2-6}&DoubleHaarNet&29.50/0.9485&25.06/0.8596&22.26/0.7630&18.92/0.5768\\
		\cline{2-6}&DoubleHaarNetPnP&\bf{30.09/0.9552}	&\bf{25.51/0.8684}&\bf{22.78/0.7773}&\bf{19.38/0.6392}\\
		\hline\multirow{5}{*}{Hill}&degraded&10.25/0.0876&8.21/0.0444&7.69/0.0286&7.47/0.0182\\
		\cline{2-6}&HaarNet&31.28/0.8541&28.06/0.7314&26.52/0.6656&24.68/0.5880\\
		\cline{2-6}&HaarNetPnP&32.81/0.9076&28.74/0.7739&27.00/0.6909&25.17/0.5973\\
		\cline{2-6}&DoubleHaarNet&33.29/0.9119&29.15/0.7836&27.15/0.6991&25.48/0.6266\\
		\cline{2-6}&DoubleHaarNetPnP&\bf{33.60/0.9187}	&\bf{29.54/0.7947}&\bf{27.51/0.6994}&\bf{25.93/0.6586}\\
		\hline\multirow{5}{*}{Man}&degraded&9.35/0.0830&7.31/0.0408&6.80/0.0266&6.56/0.0166\\
		\cline{2-6}&HaarNet&31.08/0.8782&27.85/0.7833&26.20/0.7242&24.30/0.6521\\
		\cline{2-6}&HaarNetPnP&32.70/0.9293&28.65/0.8204&26.77/0.7391&24.76/0.6733\\
		\cline{2-6}&DoubleHaarNet&33.00/0.9303&28.99/0.8295&26.93/0.7588&25.03/0.6890\\
		\cline{2-6}&DoubleHaarNetPnP&\bf{33.37/0.9375}&\bf{29.38/0.8369}&\bf{27.12/0.7670}&\bf{25.27/0.6903}\\
		\hline
	\end{tabular}
\end{table}

\begin{figure}[htbp]
	\begin{minipage}{0.3\linewidth}
		\centering
		\centerline{\includegraphics[width=1.6in]{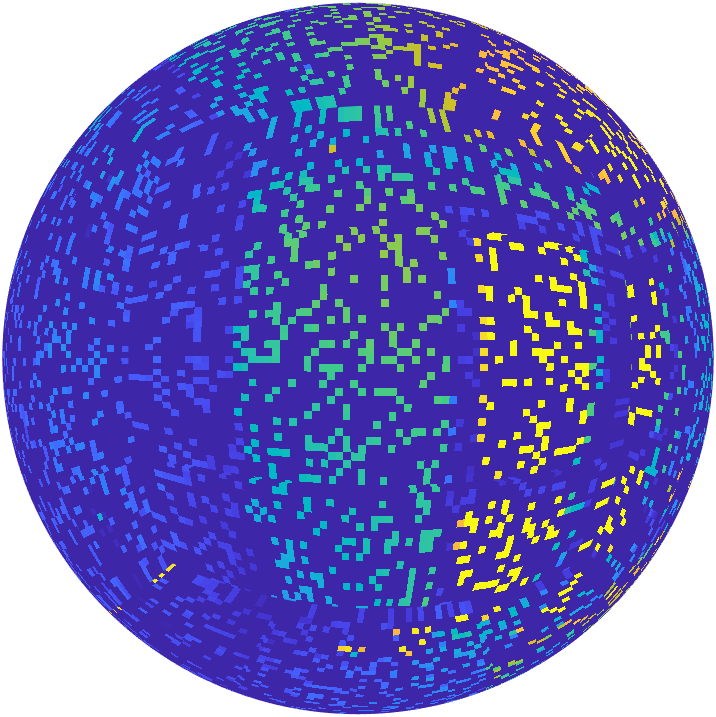}%
		}
		\centerline{\small{(a) degraded}}
		\centerline{\small{7.21/0.0612}}
	\end{minipage}
	\begin{minipage}{0.3\linewidth}
		\centering
		\centerline{\includegraphics[width=1.6in]{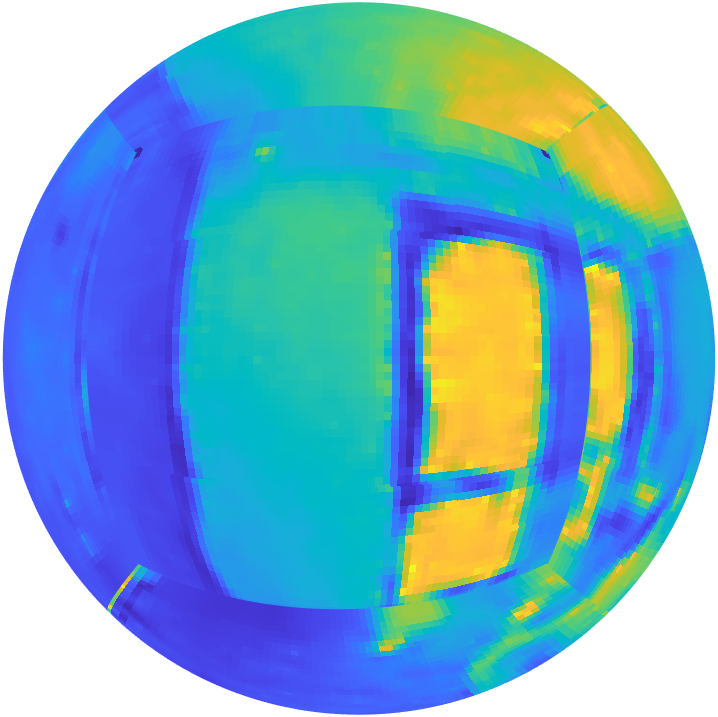}}
		\centerline{\small{(b) HaarNet}}%
		\centerline{\small{21.78/0.7525}}
	\end{minipage}
	\begin{minipage}{0.3\linewidth}
		\centering
		\centerline{\includegraphics[width=1.6in]{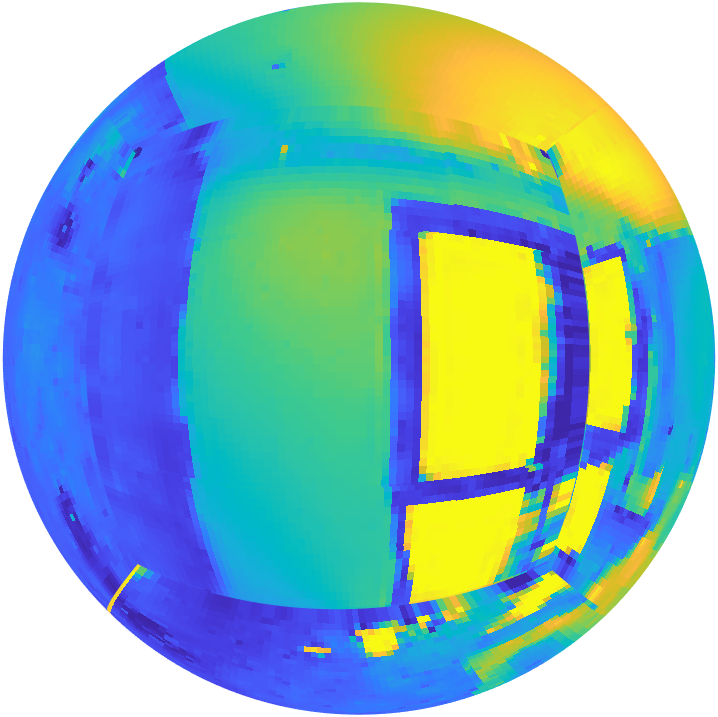}}
		\centerline{\small{(c) HaarNetPnP}}%
		\centerline{\small{22.33/0.8095}}
	\end{minipage}
	\vspace{0.1in}
	
	\begin{minipage}{0.3\linewidth}
		\centering
		\centerline{\includegraphics[width=1.6in]{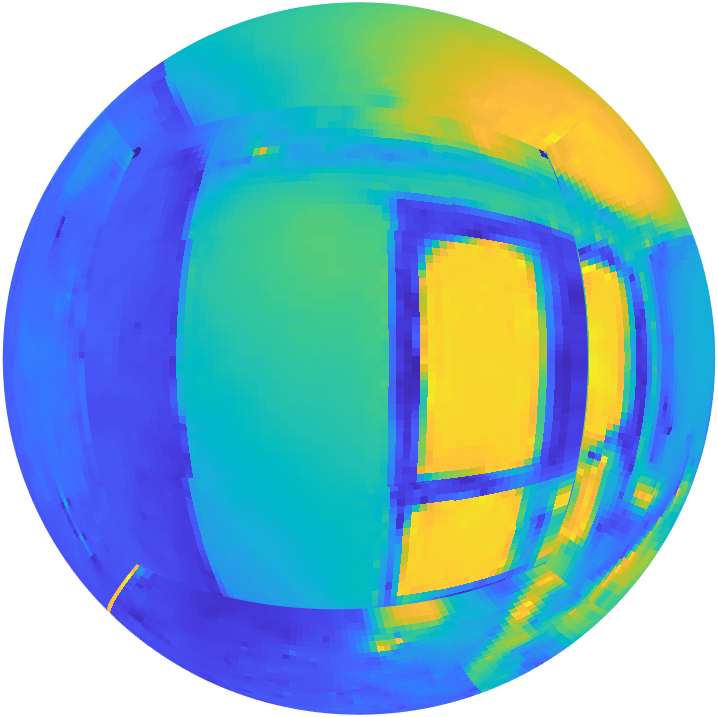}}
		\centerline{\small{(d) DoubleHaarNet}}
		\centerline{\small{23.38/0.8227}}
	\end{minipage}
	\begin{minipage}{0.3\linewidth}
		\centering
		\centerline{\includegraphics[width=1.6in]{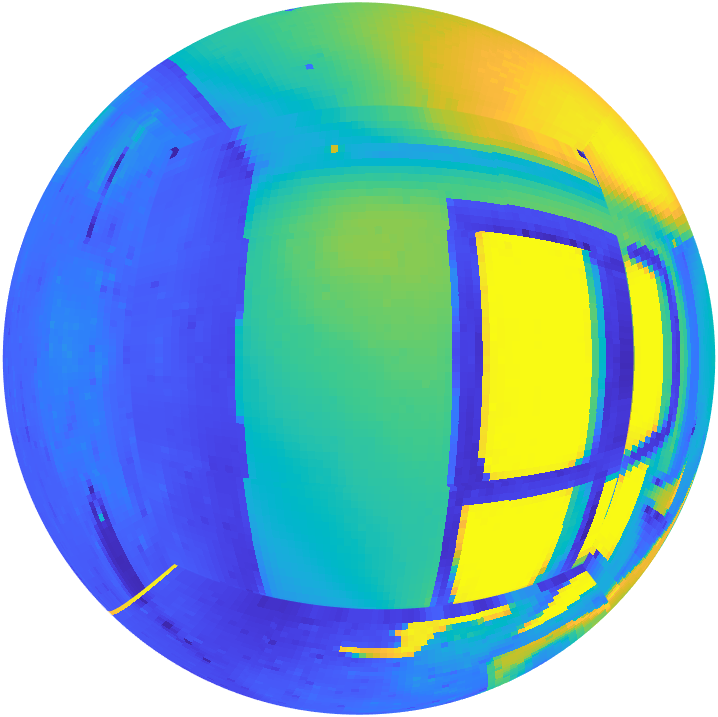}}
		\centerline{\small{(e) DoubleHaarNetPnP}}
		\centerline{\small{23.68/0.8402}}
	\end{minipage}
	\begin{minipage}{0.3\linewidth}
		\centering
		\centerline{\includegraphics[width=1.6in]{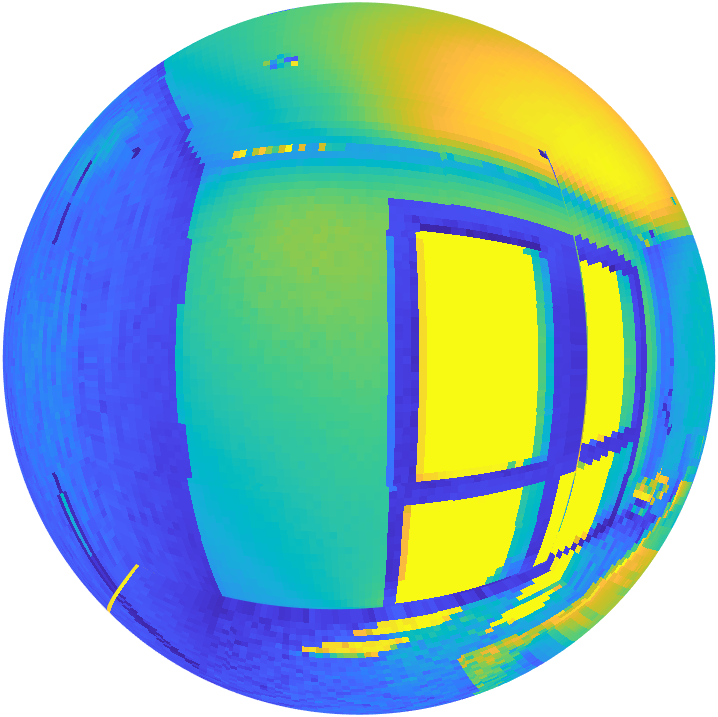}}
		\centerline{\small{(f) Ground-truth}}
		\centerline{\small{infinity/1}}
	\end{minipage}
	\caption{{The inpainting results (PSNR (dB)/SSIM) with random missing ratio 80\%. (a) the degraded image; the recovered results of (b) Haar network only; (c) plug and play with Haar network; (d) DoubleHaar network only; (e) plug and play with DoubleHaar network; (f) the original image.}}\label{f5}
\end{figure}

\begin{figure}[htbp]
	\begin{minipage}{0.3\linewidth}
		\centering
		\centerline{\includegraphics[width=1.6in]{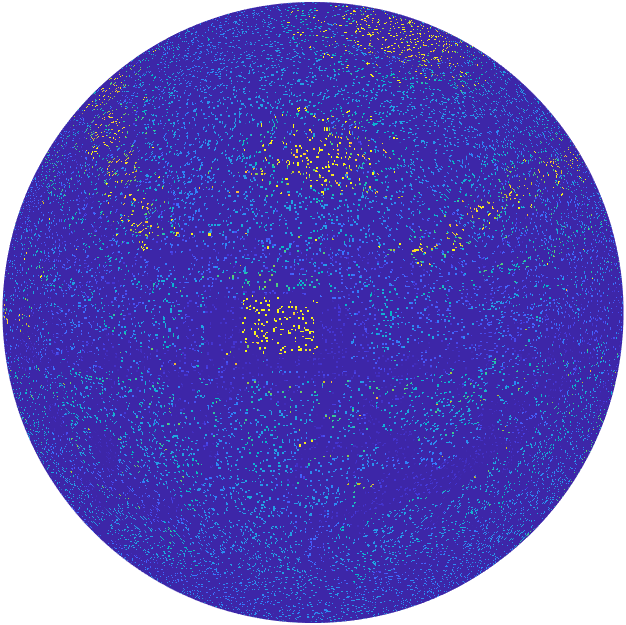}
		}
		\centerline{\small{(a) degraded}}
		\centerline{\small{7.77/0.0284}}	
	\end{minipage}
	\begin{minipage}{0.3\linewidth}
		\centering
		\centerline{\includegraphics[width=1.6in]{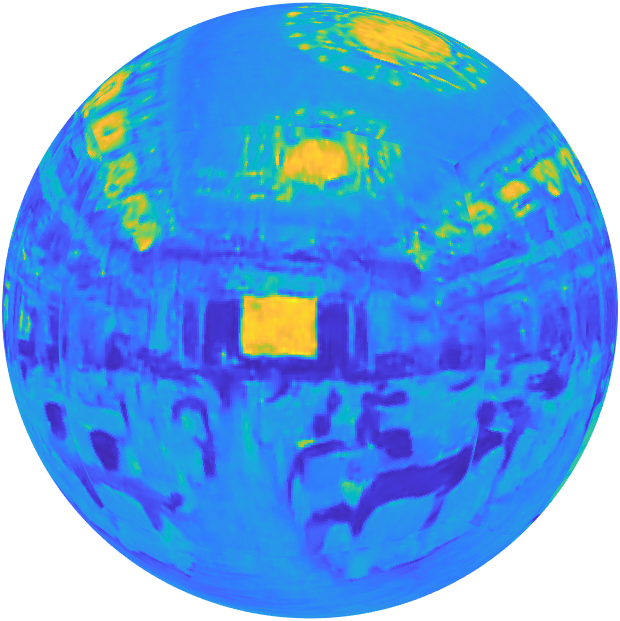}}
		\centerline{\small{(b) HaarNet}}
		\centerline{\small{19.12/0.5090}}
	\end{minipage}
	\begin{minipage}{0.3\linewidth}
		\centering
		\centerline{\includegraphics[width=1.6in]{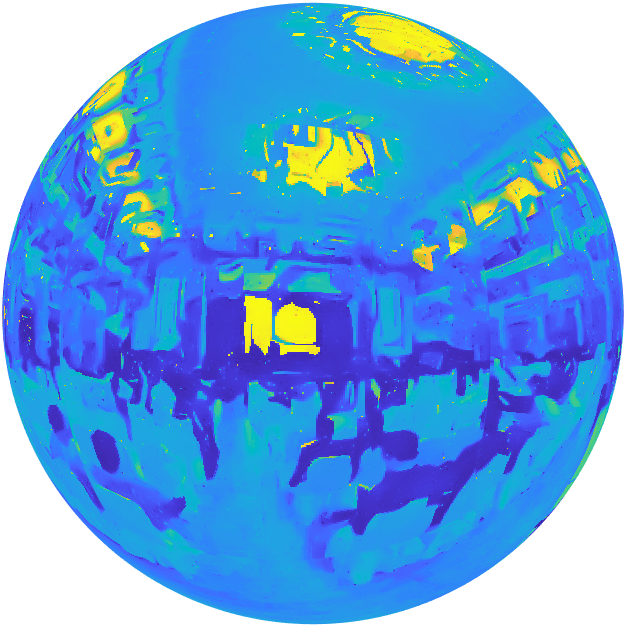}}
		\centerline{\small{(c) HaarNetPnP}}
		\centerline{\small{19.34/0.5303}}
	\end{minipage}
	\vspace{0.1in}
	
	\begin{minipage}{0.3\linewidth}
		\centering
		\centerline{\includegraphics[width=1.6in]{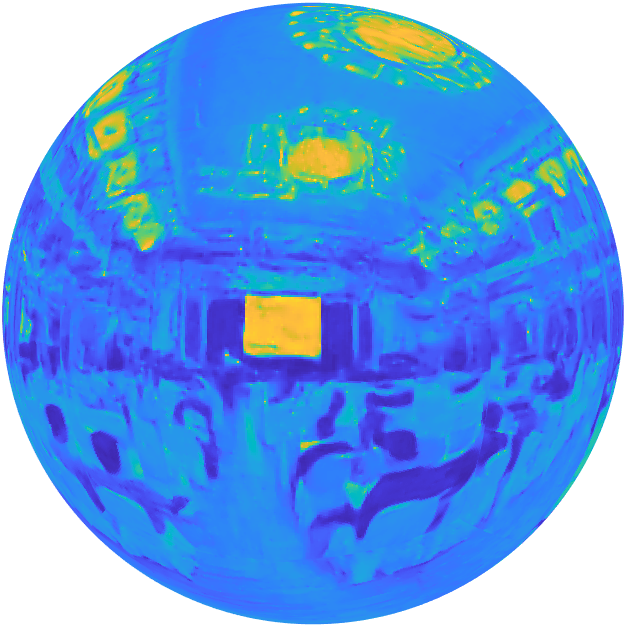}}
		\centerline{\small{(d) DoubleHaarNet}}
		\centerline{\small{19.40/0.5446}}
	\end{minipage}
	\begin{minipage}{0.3\linewidth}
		\centering
		\centerline{\includegraphics[width=1.6in]{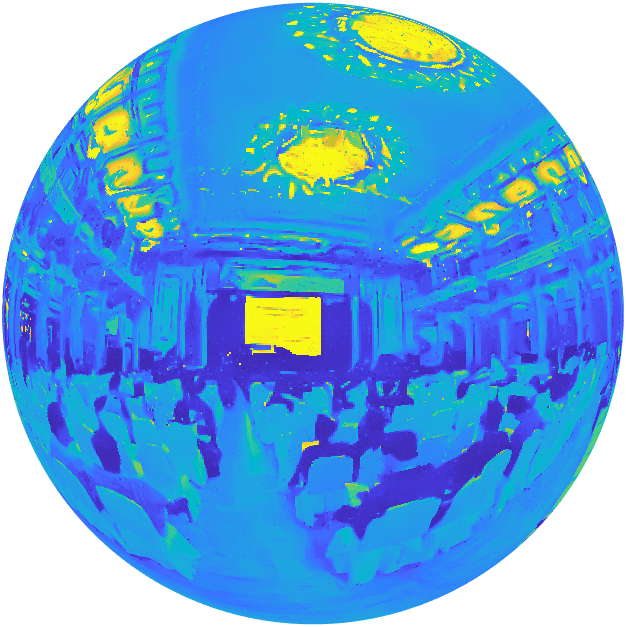}}
		\centerline{\small{(e) DoubleHaarNetPnP}}
		\centerline{\small{19.68/0.6557}}
	\end{minipage}
	\begin{minipage}{0.3\linewidth}
		\centering
		\centerline{\includegraphics[width=1.6in]{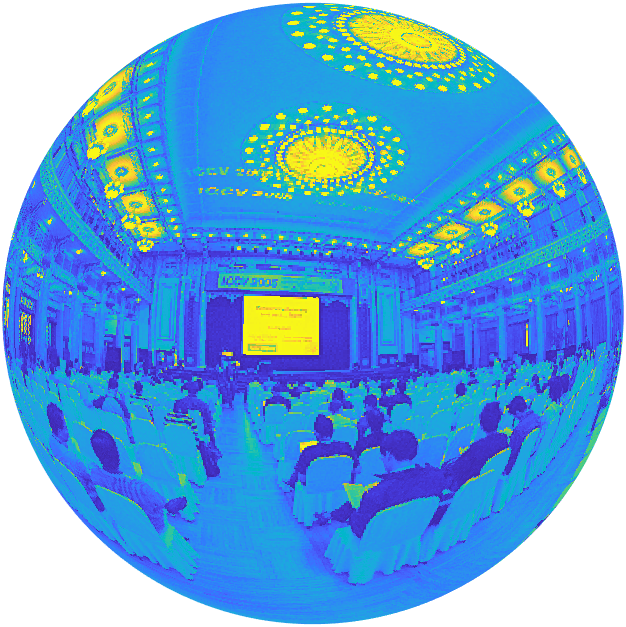}}
		\centerline{\small{(f) Ground-truth}}
		\centerline{\small{infinity/1}}
	\end{minipage}
	\caption{The inpainting results (PSNR (dB)/SSIM) with random missing ratio 90\%. (a) the degraded image; the recovered results of (b) Haar network only; (c) plug and play with Haar network; (d) DoubleHaar network only; (e) plug and play with DoubleHaar network; (f) the original image.}\label{f6}
\end{figure}

\begin{figure}[htbp]
	\begin{minipage}{0.3\linewidth}
		\centering
		\centerline{\includegraphics[width=1.6in]{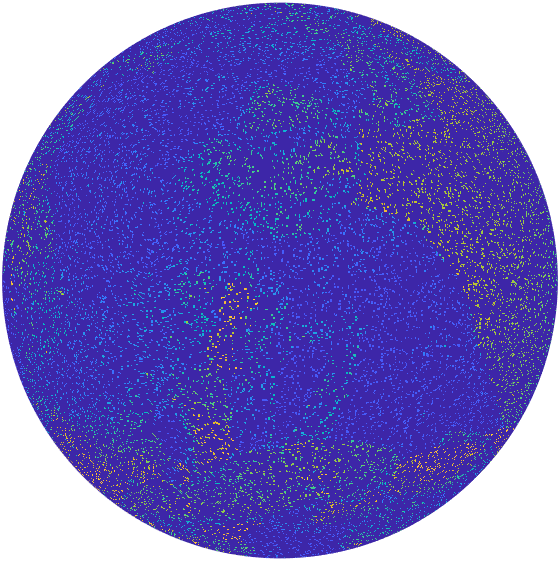}
		}
		\centerline{\small{(a) degraded}}
		\centerline{\small{6.56/0.0166}}	
	\end{minipage}
	\begin{minipage}{0.3\linewidth}
		\centering
		\centerline{\includegraphics[width=1.6in]{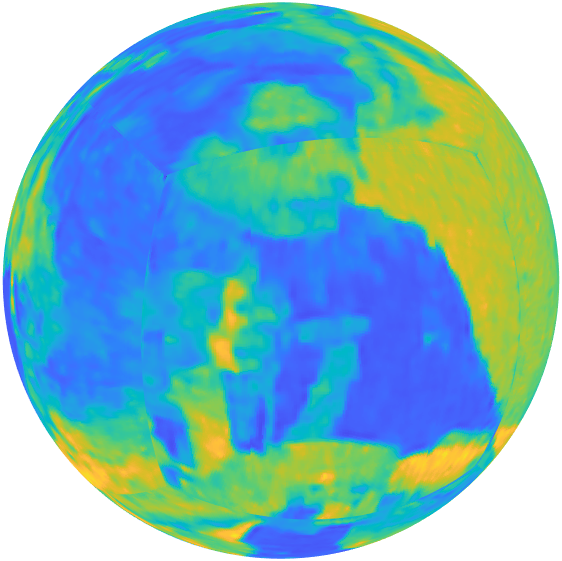}}
		\centerline{\small{(b) HaarNet}}
		\centerline{\small{24.30/0.6521}}
	\end{minipage}
	\begin{minipage}{0.3\linewidth}
		\centering
		\centerline{\includegraphics[width=1.6in]{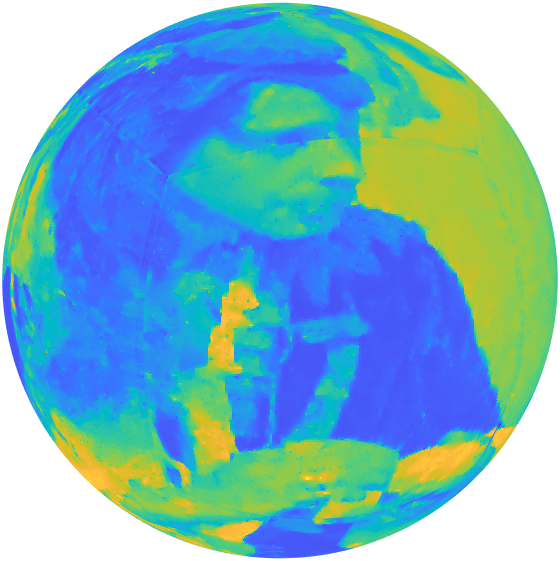}}
		\centerline{\small{(c) HaarNetPnP}}
		\centerline{\small{24.76/0.6733}}
	\end{minipage}
	\vspace{0.1in}
	
	\begin{minipage}{0.3\linewidth}
		\centering
		\centerline{\includegraphics[width=1.6in]{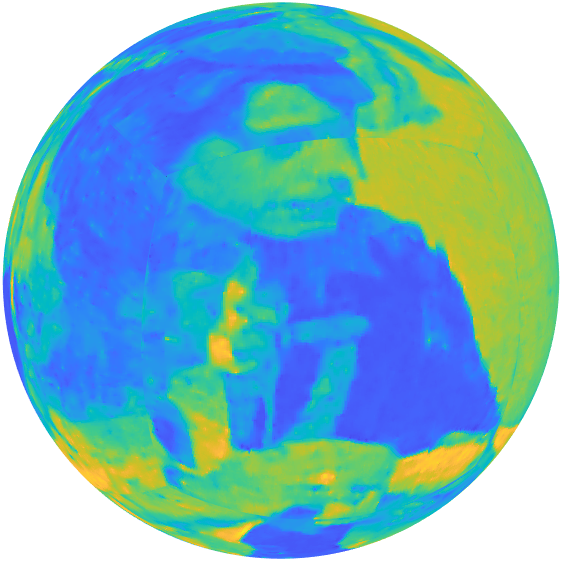}}
		\centerline{\small{(d) DoubleHaarNet}}
		\centerline{\small{25.03/0.6890}}
	\end{minipage}
	\begin{minipage}{0.3\linewidth}
		\centering
		\centerline{\includegraphics[width=1.6in]{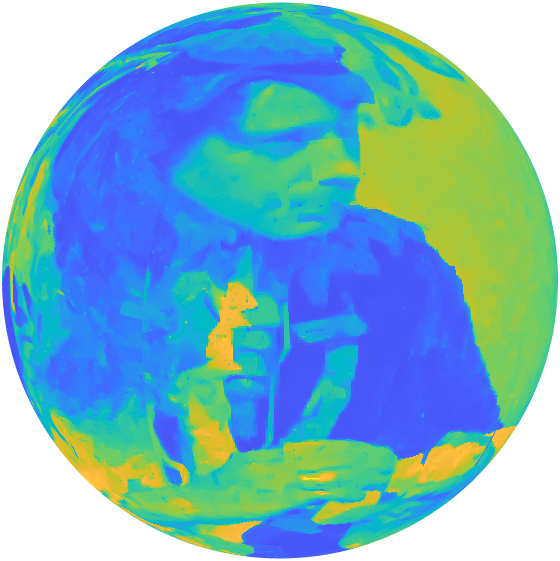}}
		\centerline{\small{(e) DoubleHaarNetPnP}}
		\centerline{\small{25.27/0.6903}}
	\end{minipage}
	\begin{minipage}{0.3\linewidth}
		\centering
		\centerline{\includegraphics[width=1.6in]{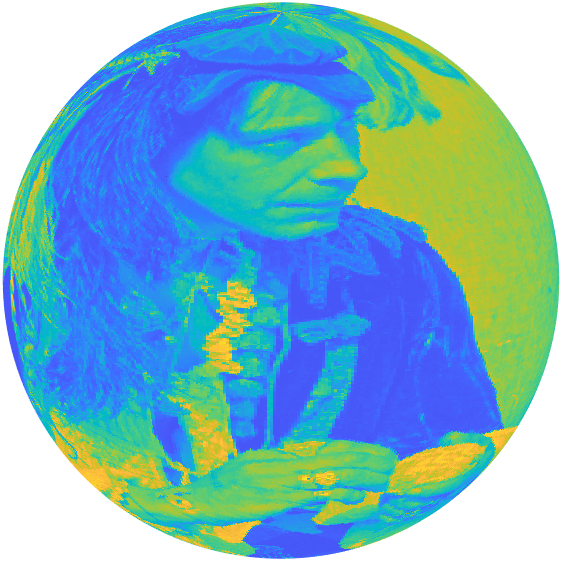}}
		\centerline{\small{(f) Ground-truth}}
		\centerline{\small{infinity/1}}
	\end{minipage}
	\caption{The inpainting results (PSNR (dB)/SSIM) with random missing ratio 95\%. (a) the degraded image; the recovered results of (b) Haar network only; (c) plug and play with Haar network; (d) DoubleHaar network only; (e) plug and play with DoubleHaar network; (f) the original image.}\label{f7}
\end{figure}

\begin{figure}[htbp]
	\hspace{0.5in}
	\begin{minipage}{0.16\linewidth}
		\centering
		\centerline{\includegraphics[width=1.2in]{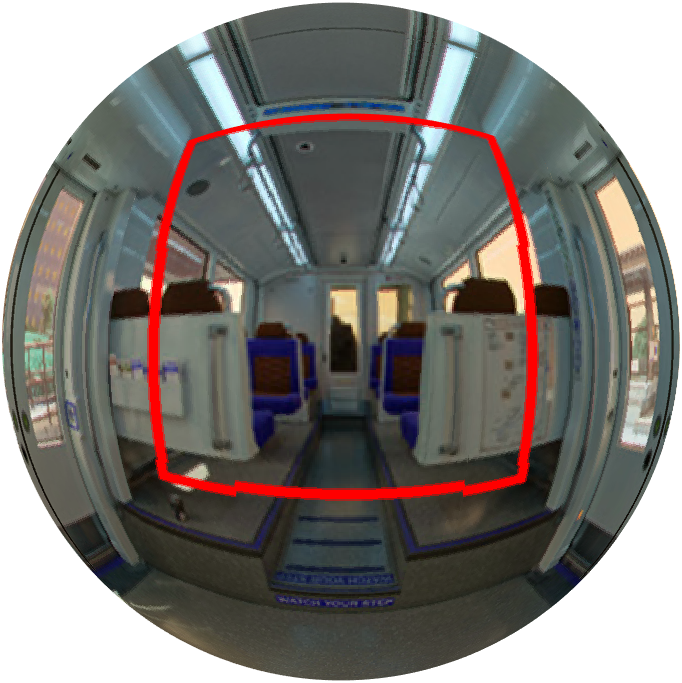}
		}
		\centerline{\small{ Color1}}
	\end{minipage}
	\hspace{0.3in}
	\begin{minipage}{0.16\linewidth}
		\centering
		\centerline{\includegraphics[width=1.2in]{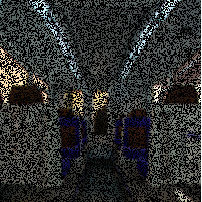}
		}
		\centerline{\small{(a) degraded}}
		\centerline{\small{8.39/0.0473}}
	\end{minipage}
	\hspace{0.3in}
	\begin{minipage}{0.12\linewidth}
		\centering
		\centerline{\includegraphics[width=1.2in]{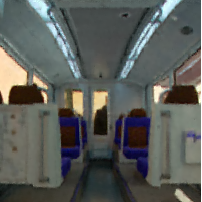}}
		\centerline{\small{(b) HaarNet}}
		\centerline{\small{31.53/0.9458}}
	\end{minipage}
	\hspace{0.3in}
	\begin{minipage}{0.16\linewidth}
		\centering
		\centerline{\includegraphics[width=1.2in]{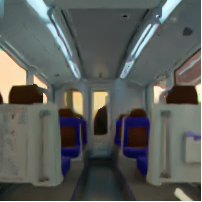}}
		\centerline{\small{(c) HaarNetPnP}}%
		\centerline{\small{32.96/0.9551}}
	\end{minipage}
	\vspace{0.1in}

	\hspace{1.86in}
	\begin{minipage}{0.16\linewidth}
		\centering
		\centerline{\includegraphics[width=1.2in]{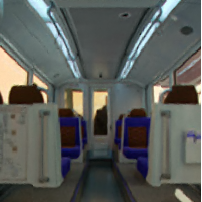}}
		\centerline{\small{(d) DoubleHaarNet}}
		\centerline{\small{36.98/0.9850}}
	\end{minipage}	\hspace{0.2in}
	\begin{minipage}{0.16\linewidth}
		\centering
		\centerline{\includegraphics[width=1.2in]{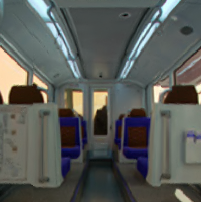}}
		\centerline{\small{(e) DoubleHaarNetPnP}}
		\centerline{\small{37.23/0.9856}}
	\end{minipage}	\hspace{0.2in}
	\begin{minipage}{0.16\linewidth}
		\centering
		\centerline{\includegraphics[width=1.2in]{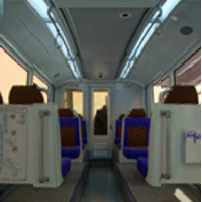}}
		\centerline{\small{(f) Ground-truth}}
		\centerline{\small{infinity/1}}
	\end{minipage}
	\caption{The inpainting results (PSNR (dB)/SSIM) with random missing ratio 50\%. (a) the degraded image; the recovered results of (b) Haar network only; (c) plug and play with Haar network; (d) DoubleHaar network only; (e) plug and play with DoubleHaar network; (f) the original image.}\label{cf5}
\end{figure}


\begin{figure}[htbp]
	\begin{minipage}{0.3\linewidth}
		\centering
		\centerline{\includegraphics[width=1.6in]{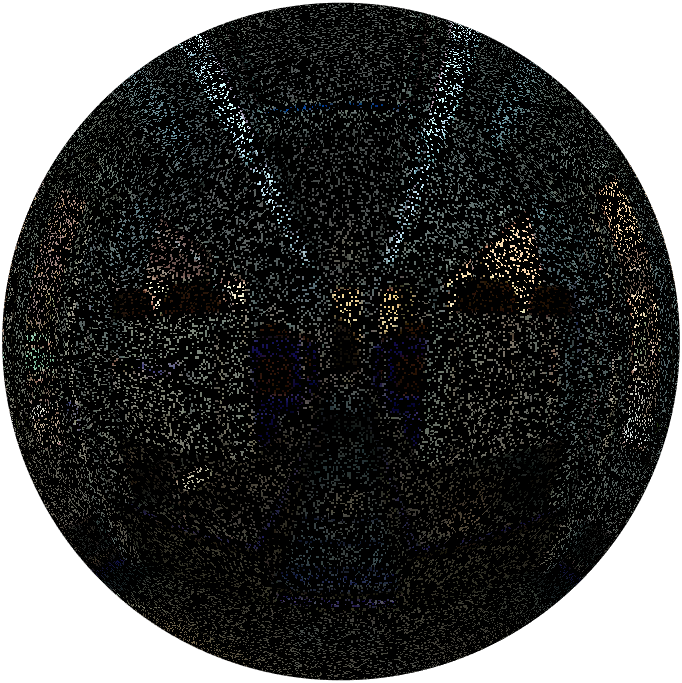}
		}
		\centerline{\small{(a) degraded}}
		\centerline{\small{8.39/0.0473}}
	\end{minipage}
	\begin{minipage}{0.3\linewidth}
		\centering
		\centerline{\includegraphics[width=1.6in]{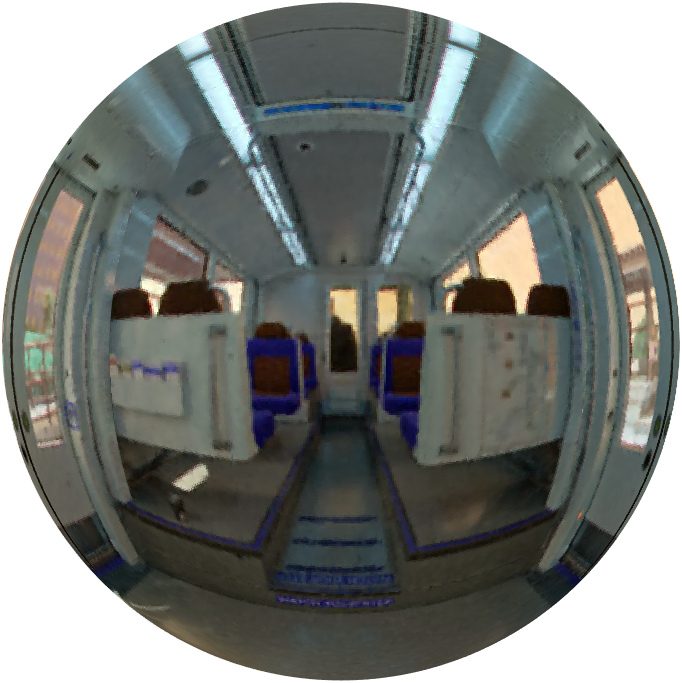}}
		\centerline{\small{(b) HaarNet}}
		\centerline{\small{31.53/0.9458}}
	\end{minipage}
	\begin{minipage}{0.3\linewidth}
		\centering
		\centerline{\includegraphics[width=1.6in]{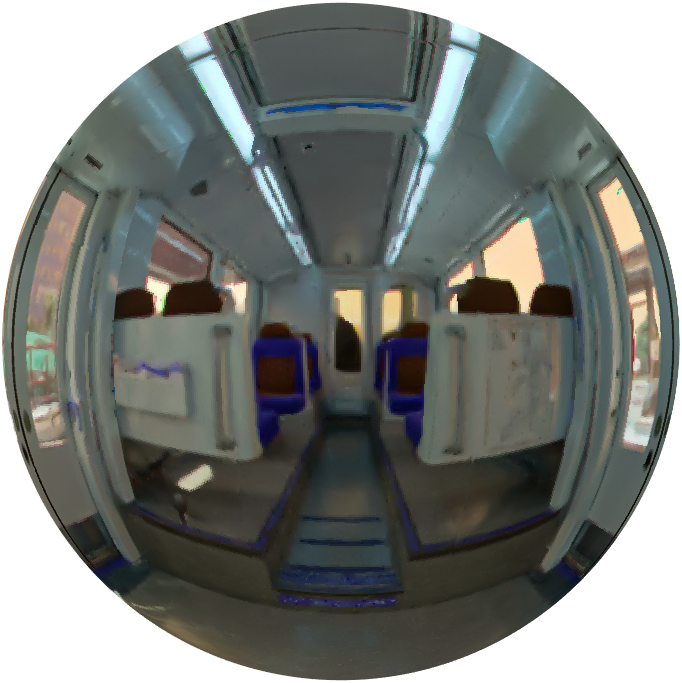}}
		\centerline{\small{(c) HaarNetPnP}}%
		\centerline{\small{32.96/0.9551}}
	\end{minipage}
	\vspace{0.1in}
	
	\begin{minipage}{0.3\linewidth}
		\centering
		\centerline{\includegraphics[width=1.6in]{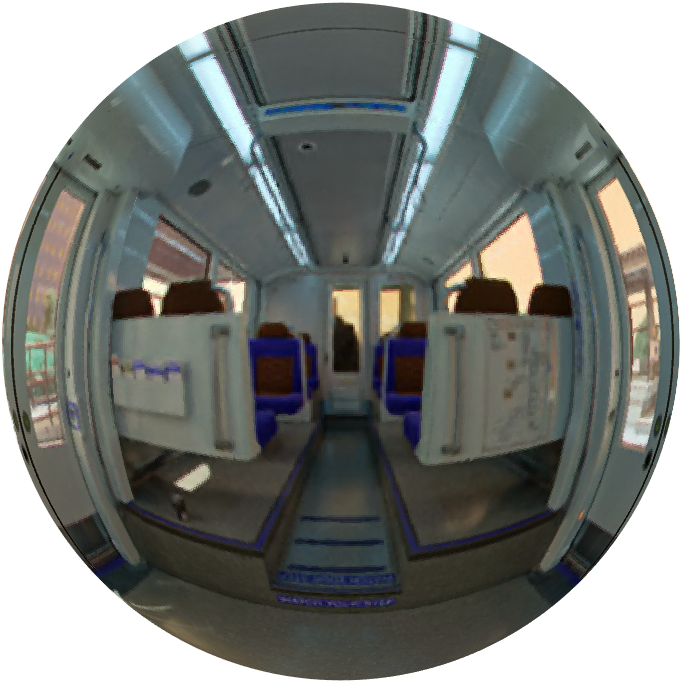}}
		\centerline{\small{(d) DoubleHaarNet}}
		\centerline{\small{36.98/0.9850}}
	\end{minipage}
	\begin{minipage}{0.3\linewidth}
		\centering
		\centerline{\includegraphics[width=1.6in]{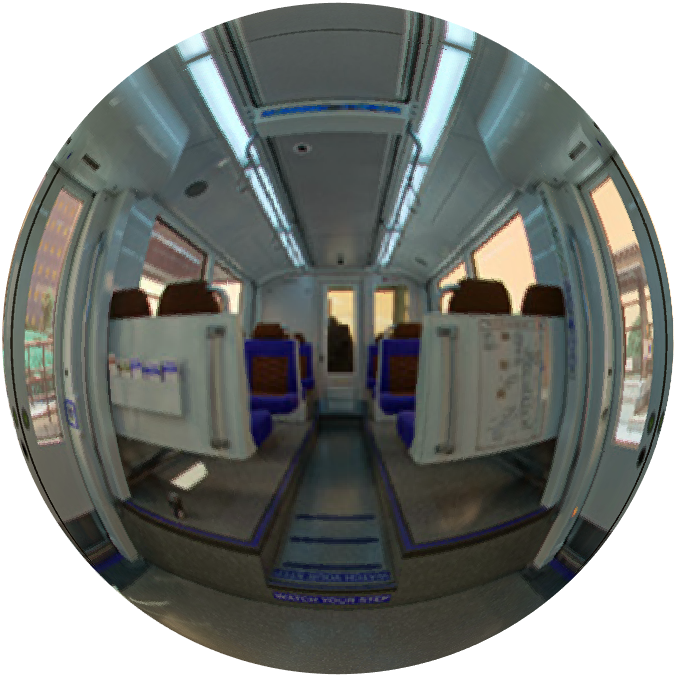}}
		\centerline{\small{(e) DoubleHaarNetPnP}}
		\centerline{\small{37.23/0.9856}}
	\end{minipage}
	\begin{minipage}{0.3\linewidth}
		\centering
		\centerline{\includegraphics[width=1.6in]{color/or00_50.png}}
		\centerline{\small{(f) Color1}}
		\centerline{\small{infinity/1}}
	\end{minipage}
	\caption{The inpainting results (PSNR (dB)/SSIM) with random missing ratio 50\%. (a) the degraded image; the recovered results of (b) Haar network only; (c) plug and play with Haar network; (d) DoubleHaar network only; (e) plug and play with DoubleHaar network; (f) the original image.}\label{f4}
\end{figure}

\begin{figure}[htbp]
	\begin{minipage}{0.3\linewidth}
		\centering
		\centerline{\includegraphics[width=1.6in]{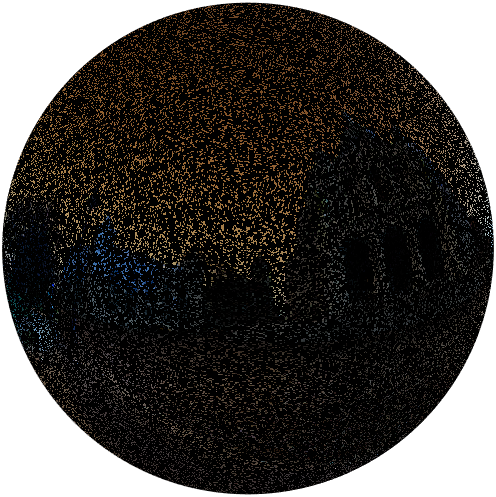}
		}
		\centerline{\small{(a) degraded}}
		\centerline{\small{8.70/0.0642}}
	\end{minipage}
	\begin{minipage}{0.3\linewidth}
		\centering
		\centerline{\includegraphics[width=1.6in]{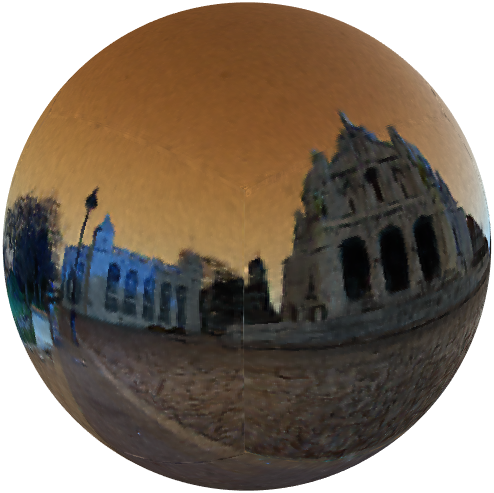}}
		\centerline{\small{(b) HaarNet}}
		\centerline{\small{24.01/0.7669}}
	\end{minipage}
	\begin{minipage}{0.3\linewidth}
		\centering
		\centerline{\includegraphics[width=1.6in]{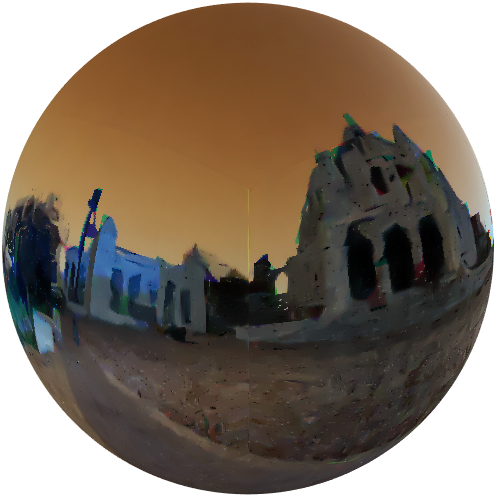}}
		\centerline{\small{(c) HaarNetPnP}}%
		\centerline{\small{25.96/0.7815}}
	\end{minipage}
	\vspace{0.1in}
	
	\begin{minipage}{0.3\linewidth}
		\centering
		\centerline{\includegraphics[width=1.6in]{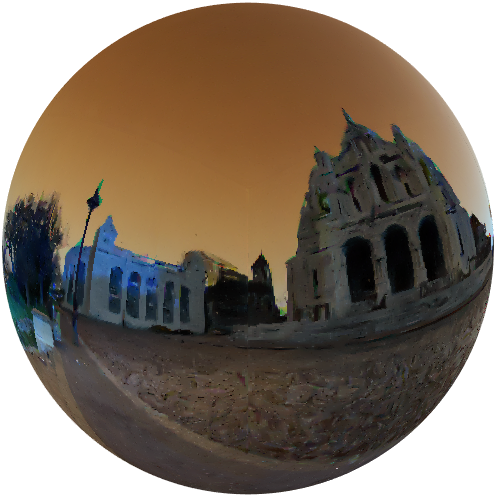}}
		\centerline{\small{(d) DoubleHaarNet}}
		\centerline{\small{26.13/0.8161}}
	\end{minipage}
	\begin{minipage}{0.3\linewidth}
		\centering
		\centerline{\includegraphics[width=1.6in]{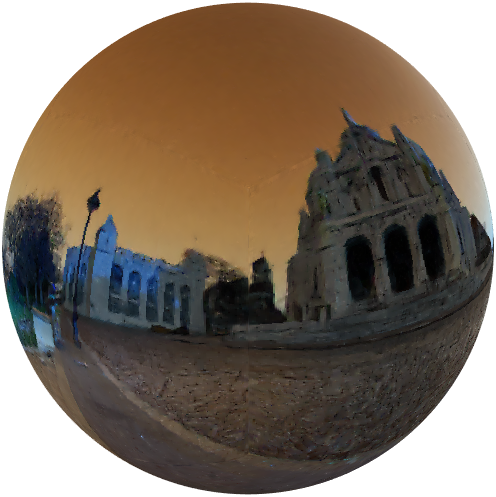}}
		\centerline{\small{(e) DoubleHaarNetPnP}}
		\centerline{\small{26.56/0.8169}}
	\end{minipage}
	\begin{minipage}{0.3\linewidth}
		\centering
		\centerline{\includegraphics[width=1.6in]{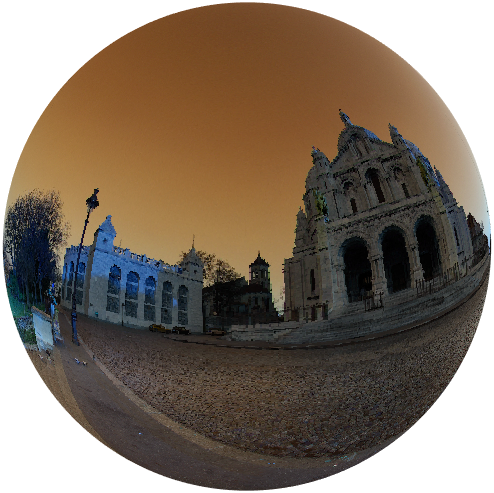}}
		\centerline{\small{(f) Color4}}
		\centerline{\small{infinity/1}}
	\end{minipage}
	\caption{The inpainting results (PSNR (dB)/SSIM) with random missing ratio 80\%. (a) the degraded image; the recovered results of (b) Haar network only; (c) plug and play with Haar network; (d) DoubleHaar network only; (e) plug and play with DoubleHaar network; (f) the original image.}\label{f4}
\end{figure}

\begin{figure}[htbp]
	\hspace{0.5in}
	\begin{minipage}{0.16\linewidth}
		\centering
		\centerline{\includegraphics[width=1.2in]{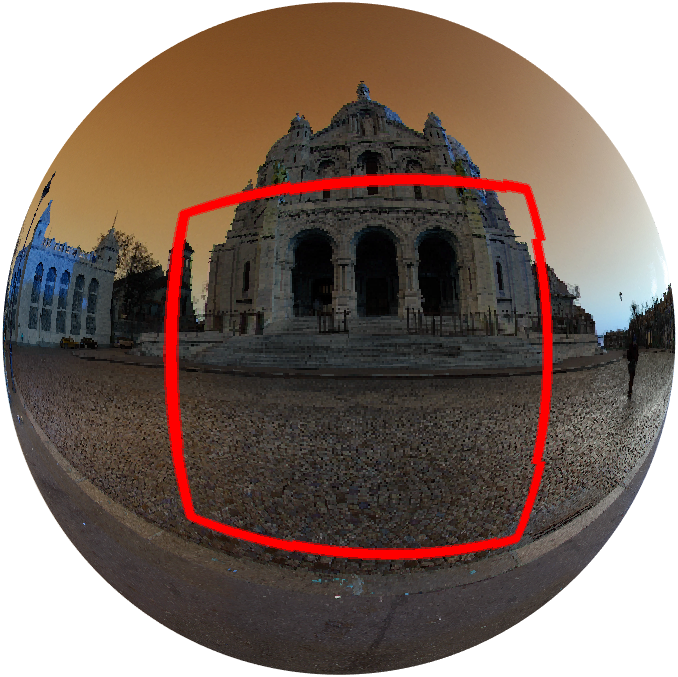}
		}
		\centerline{\small{ Color4}}
	\end{minipage}
	\hspace{0.3in}
	\begin{minipage}{0.16\linewidth}
		\centering
		\centerline{\includegraphics[width=1.2in]{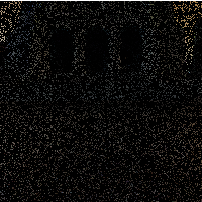}
		}
		\centerline{\small{(a) degraded}}
		\centerline{\small{8.70/0.0642}}
	\end{minipage}
	\hspace{0.3in}
	\begin{minipage}{0.12\linewidth}
		\centering
		\centerline{\includegraphics[width=1.2in]{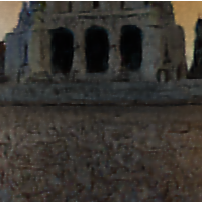}}
		\centerline{\small{(b) HaarNet}}
		\centerline{\small{24.01/0.7669}}
	\end{minipage}
	\hspace{0.3in}
	\begin{minipage}{0.16\linewidth}
		\centering
		\centerline{\includegraphics[width=1.2in]{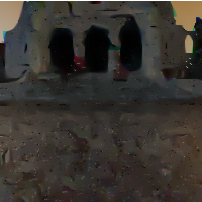}}
		\centerline{\small{(c) HaarNetPnP}}%
		\centerline{\small{25.96/0.7815}}
	\end{minipage}
	\vspace{0.1in}

	\hspace{1.86in}
	\begin{minipage}{0.16\linewidth}
		\centering
		\centerline{\includegraphics[width=1.2in]{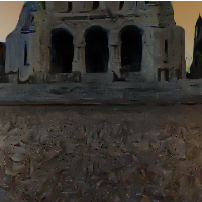}}
		\centerline{\small{(d) DoubleHaarNet}}
		\centerline{\small{26.13/0.8161}}
	\end{minipage}	\hspace{0.2in}
	\begin{minipage}{0.16\linewidth}
		\centering
		\centerline{\includegraphics[width=1.2in]{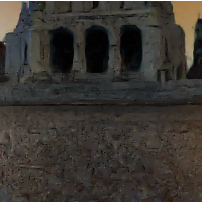}}
		\centerline{\small{(e) DoubleHaarNetPnP}}
		\centerline{\small{26.56/0.8169}}
	\end{minipage}	\hspace{0.2in}
	\begin{minipage}{0.16\linewidth}
		\centering
		\centerline{\includegraphics[width=1.2in]{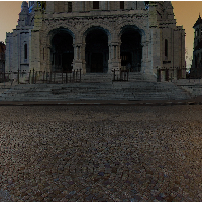}}
		\centerline{\small{(f) Ground-truth}}
		\centerline{\small{infinity/1}}
	\end{minipage}
	\caption{The color image inpainting results (PSNR (dB)/SSIM) with random missing ratio 80\%. Zoomed part of (a) the degraded image; the recovered results of (b) Haar network only; (c) plug and play with Haar network; (d) DoubleHaar network only; (e) plug and play with DoubleHaar network; (f) the original image.}\label{cf8}
\end{figure}

\begin{figure}[htbp]
	\hspace{0.5in}
	\begin{minipage}{0.16\linewidth}
		\centering
		\centerline{\includegraphics[width=1.2in]{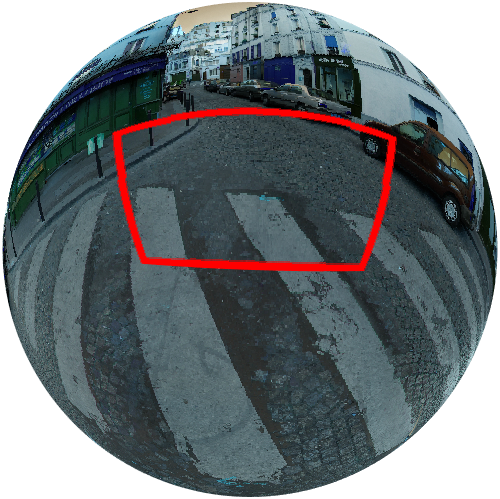}
		}
		\centerline{\small{ Color12}}
	\end{minipage}
	\hspace{0.3in}
	\begin{minipage}{0.16\linewidth}
		\centering
		\centerline{\includegraphics[width=1.2in]{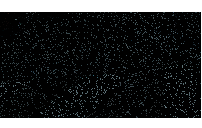}
		}
		\centerline{\small{(a) degraded}}
		\centerline{\small{6.88/0.0318}}
	\end{minipage}
	\hspace{0.3in}
	\begin{minipage}{0.12\linewidth}
		\centering
		\centerline{\includegraphics[width=1.2in]{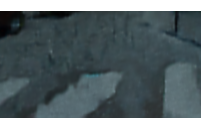}}
		\centerline{\small{(b) HaarNet}}
		\centerline{\small{21.55/0.7226}}
	\end{minipage}
	\hspace{0.3in}
	\begin{minipage}{0.16\linewidth}
		\centering
		\centerline{\includegraphics[width=1.2in]{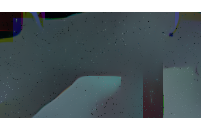}}
		\centerline{\small{(c) HaarNetPnP}}%
		\centerline{\small{21.89/0.7251}}
	\end{minipage}

	\hspace{1.86in}
	\begin{minipage}{0.16\linewidth}
		\centering
		\centerline{\includegraphics[width=1.2in]{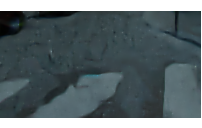}}
		\centerline{\small{(d) DoubleHaarNet}}
		\centerline{\small{21.9763/0.7485}}
	\end{minipage}	\hspace{0.2in}
	\begin{minipage}{0.16\linewidth}
		\centering
		\centerline{\includegraphics[width=1.2in]{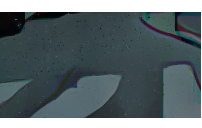}}
		\centerline{\small{(e) DoubleHaarNetPnP}}
		\centerline{\small{22.27/0.7666}}
	\end{minipage}	\hspace{0.2in}
	\begin{minipage}{0.16\linewidth}
		\centering
		\centerline{\includegraphics[width=1.2in]{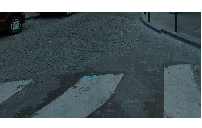}}
		\centerline{\small{(f) Ground-truth}}
		\centerline{\small{infinity/1}}
	\end{minipage}
	\caption{The color image inpainting results (PSNR (dB)/SSIM) with random missing ratio 90\%. Zoomed part of (a) the degraded image; the recovered results of (b) Haar network only; (c) plug and play with Haar network; (d) DoubleHaar network only; (e) plug and play with DoubleHaar network; (f) the original image.}\label{cf90}
\end{figure}

\begin{figure}[htbp]
	\hspace{0.5in}
	\begin{minipage}{0.16\linewidth}
		\centering
		\centerline{\includegraphics[width=1.2in]{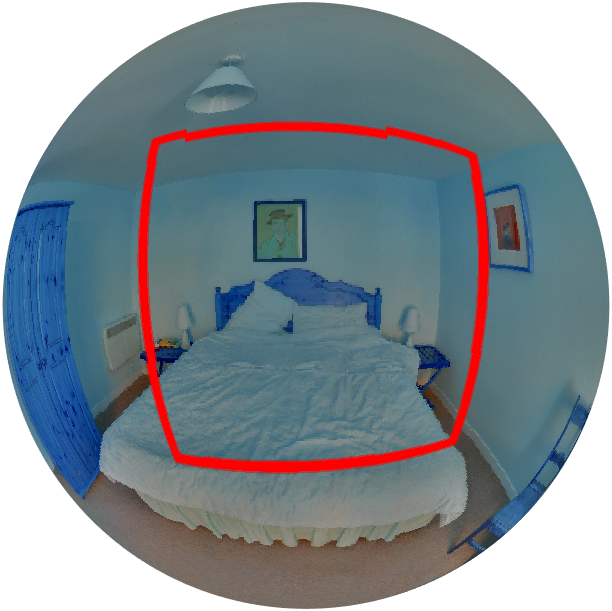}
		}
		\centerline{\small{ Color6}}
	\end{minipage}
	\hspace{0.3in}
	\begin{minipage}{0.16\linewidth}
		\centering
		\centerline{\includegraphics[width=1.2in]{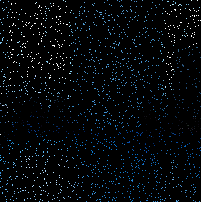}
		}
		\centerline{\small{(a) degraded}}
		\centerline{\small{7.27/0.0108}}
	\end{minipage}
	\hspace{0.3in}
	\begin{minipage}{0.12\linewidth}
		\centering
		\centerline{\includegraphics[width=1.2in]{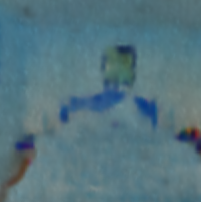}}
		\centerline{\small{(b) HaarNet}}
		\centerline{\small{27.72/0.9042}}
	\end{minipage}
	\hspace{0.3in}
	\begin{minipage}{0.16\linewidth}
		\centering
		\centerline{\includegraphics[width=1.2in]{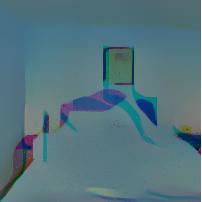}}
		\centerline{\small{(c) HaarNetPnP}}%
		\centerline{\small{28.08/0.9056}}
	\end{minipage}
	\vspace{0.1in}

	\hspace{1.86in}
	\begin{minipage}{0.16\linewidth}
		\centering
		\centerline{\includegraphics[width=1.2in]{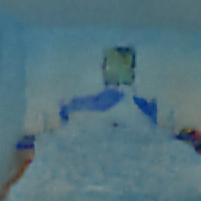}}
		\centerline{\small{(d) DoubleHaarNet}}
		\centerline{\small{28.84/0.9193}}
	\end{minipage}	\hspace{0.2in}
	\begin{minipage}{0.16\linewidth}
		\centering
		\centerline{\includegraphics[width=1.2in]{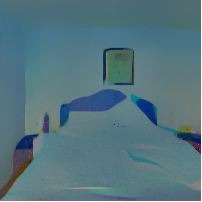}}
		\centerline{\small{(e) DoubleHaarNetPnP}}
		\centerline{\small{29.09/0.9216}}
	\end{minipage}	\hspace{0.2in}
	\begin{minipage}{0.16\linewidth}
		\centering
		\centerline{\includegraphics[width=1.2in]{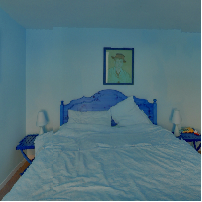}}
		\centerline{\small{(f) Ground-truth}}
		\centerline{\small{infinity/1}}
	\end{minipage}
	\caption{The color image inpainting results (PSNR (dB)/SSIM) with random missing ratio 95\%. Zoomed part of (a) the degraded image; the recovered results of (b) Haar network only; (c) plug and play with Haar network; (d) DoubleHaar network only; (e) plug and play with DoubleHaar network; (f) the original image.}\label{cf95}
\end{figure}

\subsection{Evaluation metrics}
\y{To demonstrate the effectiveness of the proposed scheme, we test the images with random missing values, i.e., the missing ratio with $50\%, 80\%, 90\%, 95\%$. With the built-in function in numpy, we use the command `numpy.random.rand' to generate random values of the same size as the input image. Then let the value which is great than the missing ratio equal to 1 and less than the missing ratio equal to 0. Then we have the missing operator, with the function `numpy.multiply' mapping the missing operator and the groundtruth image to the observed image. 
The peak signal-to-noise ratio (PSNR) and structural similarity index (SSIM) are used to evaluate the performance of the inpainting results. To be specific, with the reference $x$ and the obtained result $x^*$, the PSNR is defined as 
\begin{equation}
\mathrm{PSNR}(x,x^*)=20 \log_{10} \frac{255}{\frac{1}{m n}\left\|x^{\star}-x\right\|},
\end{equation}
where $\Vert\cdot\Vert$ denotes a Frobenius norm.
The SSIM is defined as 
\begin{equation}
\operatorname{SSIM}\left(x, x^{\star}\right)=\frac{\left(2 \mu_{x} \mu_{x^{\star}}+C_{1}\right)\left(2 \sigma_{x x^{\star}}+C_{2}\right)}{\left(\mu_{x}^{2}+\mu_{x^{\star}}^{2}+C_{1}\right)\left(\sigma_{x}^{2}+\sigma_{x^{\star}}^{2}+C_{2}\right)}
\end{equation}
where $\mu_{x}$, $\mu_{x^*}$ and $\sigma_{x}$, $\sigma_{x^*}$, $\sigma_{xx^*}$ are the mean and the standard deviation of $x$ and $x^*$, respectively. The positive constants $C_1$ and $C_2$ are used to avoid a null denominator, which are defaulted by the build-in ssim function. 
}


\subsection{Results}
We make a detailed comparison of our method. More specifically, the methods with single net (S2HaarNet and DoubleS2haarNet) and the methods with plug-and-play (S2HaarNetPnP and DoubleS2haarNetPnP) are compared.  
First of all, the three datasets are tested. The results of datasets `F6', `F7', and `F8' are listed in Table \ref{t1}, from which we know that the proposed methods are robust in different degradation. For example, when the missing ratio is from $50\%$ up to $95\%$, our methods always have competitive restoration results. Besides, the results based on the plug-and-play approach are better than the one with only CNN, which also illustrate the effectiveness of the proposed plug-and-play scheme. Moreover, the testing of two different datasets and different combinations of the proposed model also gives a strong validity to our scheme.  
On the other hand, five grey images  are also tested in this paper. We list the numerical results in Table \ref{t2}, from which the plug-and-play-based models also have better restoration results. 

The visual results are present in Figures~\ref{f4}--\ref{f7} with missing ratio $50\%, 80\%, 90\%, 95\%$, respectively. Figure~\ref{f4} (a) is the observed image with a low missing ratio ($50\%$). As we can see from the results (b)--(e), most of the objects in the image are recovered. However, with detailed observation, we know that the plug-and-play-based methods have more competitive performance. 
With the missing ratio up to $80\%$, there are some outlines of the original image that can be seen in Figure \ref{f5} (a). It turns out that the results of the HaarNet and DoubleHaarNet have quite satisfactory results. As the plug-and-play is applied in (b) and (d) respectively, the results of inpainting are greatly improved. For the low sample ratio, from Figure \ref{f6} (a) and Figure \ref{f7} (a), the details of the image are almost disappeared. With this low observation, our models also can recover the images with good quality. The above visual results demonstrate again the robustness and effectiveness of the proposed inpainting models.

The experiments on color images are also conducted in Figures~\ref{cf5}--\ref{cf95}, which illustrate the good generalization of our model to color images.
Note that our results are slightly over-smoothed for lower sample rates, such as 95$\%$. Figure \ref{cf95} shows that, in contrast to other results, our solutions can better restore the structures of images, which is consistent with the results of the majority of plug-and-play-based works.

\section{Conclusion and further remarks}
\label{sec:remarks}
In this work, we presented the doubleHaarNetPnP model for image inpainting. We remark that (a) low-rank framelet coefficient regularizer is introduced to learn, (b) a new denoiser DoubleHaarNet for spherical image inpainting is proposed and in the experiments, it is powerful for inpainting task with the deep Plug-and-play framework, and (c) Directional spherical Haar framelet is employed to capture directional texture information to enhance the learning ability of the model and the neural network. Experiments evaluated on various images illustrate the effectiveness of the proposed method for spherical image inpainting problems. Since our framework is general and flexible, the corresponding model for other spherical imaging tasks for example spherical image segmentation.   


\section*{Acknowledgement}
This work was supported in part by the National Key R\&D Program of China under Grant 2021YFE0203700, Grant NSFC/RGC N\_CUHK 415/19, Grant NSFC Nos. 11871210, 11971215, and 61971292, Grant ITF MHP/038/20, Grant RGC 14300219, 14302920, 14301121, and CUHK Direct Grant for Research. And in part by Hong Kong Research Grant Council GRF 12300218, 12300519, 17201020, 17300021, C1013-21GF, C7004-21GF and Joint NSFC-RGC N-HKU76921. And in part by HKRGC Grants Nos. CUHK14301718, CityU11301120, C1013-21GF, CityU Grant 9380101.

\bibliographystyle{plain}


\end{document}